\documentclass[useAMS,usenatbib]{mn2e}
\usepackage{latexsym,graphicx,natbib,amssymb,array,aas_macros}

\newcommand{\msun}{M_{\odot}}
\newcommand{\myr}{\msun\;{\rm yr}^{-1}}
\newcommand{\mpc}{M_{\odot}\;{\rm pc}^{-3}}

\newcommand{\fb}{f_{\rm b}}

\newcommand{\meclmin}{M_{\rm ecl,min}}

\newcommand{\al}{\textsc{AstraLux}}
\newcommand{\minms}{\textsc{MinMS}}

\newcommand{\rh}{r_h}

\title[M-dwarfs as tracers of star and BD formation]{M-dwarf binaries as tracers of star and brown dwarf formation}

\author[M. Marks et. al.] { Michael Marks,$^{1,2}$\thanks{e-mail: mmarks@astro.uni-bonn.de (MM)}, Markus Janson$^3$, Pavel Kroupa$^1$, Nathan Leigh$^{4,5}$ and Ingo Thies$^1$\\
$^1$Helmholtz-Institut f\"ur Strahlen- und Kernphysik, University of Bonn, Nussallee 14-16, D-53115 Bonn\\
$^2$Clara-Fey-Gymnasium, Rheinallee 5, D-53173 Bonn\\
$^3$Department of Astronomy, Stockholm University, AlbaNova University Center, SE-106 91 Stockholm\\
$^4$Department of Astrophysics, American Museum of Natural History, Central Park West and 79th Street, NY 10024, USA\\
$^5$Department of Physics, University of Alberta, CCIS 4-183, Edmonton, AB, T6G 2E1, Canada}

\begin{document}

\date{Accepted ????. Received ?????; in original form ?????}

\pagerange{\pageref{firstpage}--\pageref{lastpage}} \pubyear{2014}

\maketitle

\label{firstpage}

\begin{abstract}
The separation distribution for M-dwarf binaries in the \al~survey is narrower and peaking at smaller separations than the distribution for solar-type binaries. This is often interpreted to mean that M-dwarfs constitute a continuous transition from brown dwarfs (BDs) to stars. Here a prediction for the M-dwarf separation distribution is presented, using a dynamical population synthesis (DPS) model in which ``star-like'' binaries with late-type primaries ($\lesssim1.5\msun$) follow universal initial distribution functions and are dynamically processed in their birth embedded clusters. A separate ``BD-like'' population has both its own distribution functions for binaries and initial mass function (IMF), which overlaps in mass with the IMF for stars. Combining these two formation modes results in a peak on top of a wider separation distribution for late M-dwarfs consistent with the late \al~sample. The DPS separation distribution for early M-dwarfs shows no such peak and is in agreement with the M-dwarfs in Multiples (\minms) data. We note that the latter survey is potentially in tension with the early \al~data. Concluding, the \al~and \minms~data are unable to unambiguously distinguish whether or not BDs are a continuous extension of the stellar IMF. Future observational efforts are needed to fully answer this interesting question. The DPS model predicts that binaries outside the sensitivity range of the \al~survey remain to be detected. For application to future data, we present a means to observationally measure the overlap of the putative BD-like branch and the stellar branch. We discuss the meaning of universal star formation and distribution functions.
\end{abstract}

\section{Introduction: Common or separate populations?}
\label{sec:intro}
Binaries are a dominant channel of star formation \citep{gk2005,dk2013rev,reipurth2014rev} which is a result of the angular-momentum problem in star formation. The distribution of their orbital separations are commonly described by bell-shaped distributions that depend on the spectral-type of the primary.

The pioneering work of \citet{DuqMay91} showed a wide distribution peaking at about $\approx30$~AU, and a binary fraction of $58\pm10$\% for G-dwarf binaries in the Galactic field. More recent analyses largely confirmed their findings \citep{r2010,tok2014fg} but suggest a somewhat lower binary fraction of $\approx46\pm2$\%. Subsequent work by \citet{fm1992m} suggested a similarly wide distribution for the M-dwarf binaries which peaks at a separation comparable to solar-type stars. Their binary fraction was estimated to be $\approx42\pm9$\%. \citet{d2004m}'s M-dwarf sample confirmed a wide distribution. Similar results are obtained for K-dwarf binaries \citep{mayor1992k}.

Studies of BD binaries later showed the distribution of their orbital separations to be significantly narrower and the binary fraction of around $\approx15$\% to be lower than that for stars \citep{close2003bd,bouy2003}. This has led to the suggestion that BDs are a population separate from the hydrogen-burning stars.

Due to the different behaviour of stars and BDs in terms of their pairing characteristics, \citet{tk2007,tk2008tau} suggested an IMF which has a discontinuity around the hydrogen-burning mass limit, but with a non-negligible overlap of both populations, i.e. some BDs may form ``star-like'' while some very low-mass (VLM) stars form ``BD-like'', i.e in a fragmenting circumstellar disc. In this model both populations pair their objects among each other to form binaries but do not mix in pairing. The two-component IMF obtained after correcting the star counts for unseen companions is consistent with observations \citep{tk2008tau} and smoothed particle hydrodynamical (SPH) computations which produce BDs from encounter-triggered pertubations in circumstellar discs \citep{thies2010bd}. The models of Thies \& Kroupa furthermore support an observed BD desert \citep{mzb2003desert,grether2006desert,dieterich2012}. Additionally, \citet{thies2015} recently demonstrated that assuming BDs to be a continuous extension to the hydrogen-burning stars leads to an overestimation of the numbers of BDs when compared with actual observations, confirming the deductions of \citet{kb2003tau}. \citet{li2015fragmentation} recently investigated the outcome of a similar disc fragmentation mode (through self-instability of the disc, though, not through encounter-induced fragmentation). By combining SPH and \textsc{Nbody} techniques, they found a narrow separation distribution for BD-like objects peaking between $5$ and $10$~AU after $10$~Myr of evolution (their fig.~11), similar to the BD Galactic field properties, and within the limits of their initial conditions, their results are in good agreement with the observed BD desert.

The distinguishing characteristics of the two populations are now being challenged by recent observations of M-dwarfs in the Galactic field. The \al~survey \citep{janson2012m,janson2014m} suggests that the M-dwarf separation distribution is significantly narrower than that of solar-type stars. It is unlikely that a solar-type separation distribution \citep{DuqMay91,r2010} is a parent function for both early and late M-dwarfs. This is interpreted as evidence for continuous star formation from BDs to stars, where a population's binary fraction increases smoothly with increasing mass of the primary component, along with the peak and width of the separation distribution \citep{dk2013rev}. \citet{pm2014} argue that these observations in combination with the multiplicity properties of BDs \citep{tk2007}, of solar-type stars \citep{r2010} and of A-type stars \citep{derosa2014vast} disfavour universal initial binary properties throughout these spectral types. Instead they consider a scenario in which the 
field binary population is indicative of the primordial multiplicity conditions in the star formation regions they were born in. This would violate the universality hypothesis\footnote{Note that the universality hypothesis of star formation refers to an environment independence and does not preclude a mass-dependence (see Sec.~\ref{sec:meaning}).} of star formation according to which the universality of the IMF and binary populations are strongly coupled \citep{k2001,kp2011}.

The influence of a separate BD-like population superposed onto a star-like population on the multiplicity properties of M-dwarfs has hitherto not been tested. This is the aim of the present research paper. In Section~\ref{sec:surveys} the characteristics of the here used surveys are summarized. Section~\ref{sec:model} outlines how the Galactic field population is modelled. The results of our analysis are given in Section~\ref{sec:results} and discussed in Section~\ref{sec:discuss}. The paper closes with a summary in Section~\ref{sec:sum}.

\section{Surveys}
\label{sec:surveys}
\begin{figure}
 \includegraphics[width=0.45\textwidth]{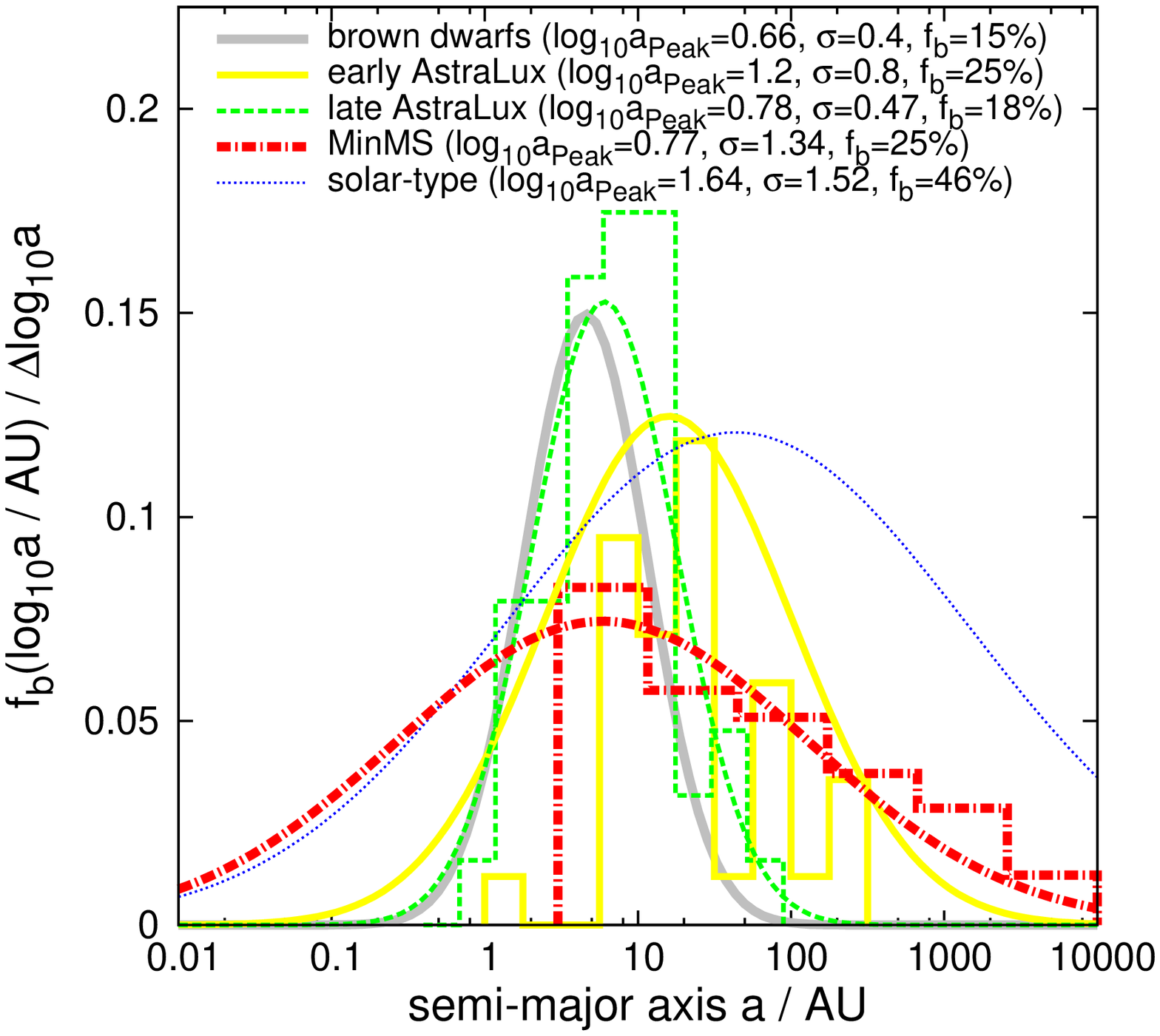}
 \caption{Observed orbital separation distributions and Gaussian fits for binaries in the Galactic field. M-dwarf data from the \al~survey \citep[histograms and corresponding same-linetype Gaussian distributions, statistically cleaned samples only;][]{janson2012m,janson2014m} and the \minms~data \citep[lower limit,][]{wardduong2015}, BD data from the VLMBA archive as described in \citet{tk2007}, and solar-type data from \citet{r2010}. The key provides the distribution peak ($\log_{10}(a_{\rm{Peak}}/\rm{AU})$), the distribution width ($\sigma$) and the binary fraction ($\fb$) within the respective sensitivity ranges as stated by the authors.}
 \label{fig:obsdist}
\end{figure}
\subsection{\al}
\label{sec:al}
In a lucky-imaging survey the \al~survey investigates binarity among ``early'' \citep[spectral types M0-M6,][]{janson2012m} and ``late'' M-dwarfs \citep[M2-M8,][]{janson2014m} in the Galactic field. Although the two samples are not distinct in mass we nevertheless henceforth refer to ``early'' and ``late'' for the sample with the smaller \citep{janson2012m} and larger \citep{janson2014m} median spectral type, respectively.

Constructing two samples distinct in spectral type or primary mass would be desirable in order to amplify the observed mass-dependency under study. The selection criteria for the two samples were however systematically different (aside from the different spectral type ranges). The early sample was selected on the basis of X-ray brightness, while the late sample was based on more nearby stars with a cut-off in infrared brightness. This means that the completeness of the surveys are systematically different, and it also follows that the early sample is systematically young and often pre-main sequence, while the late sample is not. This in turn means that the stellar masses and the spectral types the different samples correspond to are systematically different as the temperature evolves with age until the main sequence. Hence, a cut in spectral type between two sub-populations in a combined sample would be inaccurate. The youth of the early sample additionally implies that masses of some targets are still subject to change which might further diminish, or amplify, the overall offset in primary mass. Furthermore, the majority of stars in the late sample have parallactic distances while the majority of stars in the early sample do not, so adding stars from the early sample to the late sample would compromise the quality of the semi-major axis distribution determination of the latter. This is why we will focus on the late-type sample whose separation distribution is more reliable. Our conclusions are not affected by investigating the two samples individually since the DPS model (Sec.~\ref{sec:model}) takes into account the individual and overlapping mass-ranges.

\citet{janson2014m} define a ``statistically cleaned'' (SC) sample.  If the combined magnitude of the binary components exceeded the limiting survey brightness, the authors removed the target from their statistical analysis. Had the primary been a single object, it would not have been selected. This SC sample then has 48 binaries among 268 targets, corresponding to a binary fraction of $17.9$\% \textit{within the sensitivity range}, $\log_{10}(a/AU)=-0.4$ to $1.95$, where $a$ is the semi-major axis. Fig.~\ref{fig:obsdist} shows that, in comparison to the separation distribution for solar-type stars \citep{r2010}, both early and late M-dwarf samples exhibit separation distributions which are rather narrow around their peak. For the two \al~samples in Fig.~\ref{fig:obsdist} we note the following two points:
\begin{enumerate}
 \item The semi-major axis distribution of late M-dwarfs is strikingly close to that for BDs.
 \item The amplitude of the peak in the late M-dwarf semi-major axis distribution is larger than the one for early M-dwarfs.
\end{enumerate}
At first glance, item~(i) appears to favour the continuity of star formation over and above the hydrogen-burning mass limit. However, it will be demonstrated that the origin of the observed proximity of both distributions might be that a separate BD-like population overlaps with a population of star-like bodies. In addition, the here devised DPS model yields item~(ii).

\subsection{\minms}
\label{sec:minms}
The 15~pc volume-limited M-dwarfs in Multiples survey \citep[\minms,][]{wardduong2015} detects K7--M6 binaries over a separation range $\log_{10}(a/AU)\approx0.5-4$, using an infrared adaptive optics (AO) technique to find close companions ($\approx10^0-10^2$~AU) and digitised wide-field archival plates for wide companions ($10^2-10^4$~AU) each covering multiple epochs. The targets stem from a reduced \textit{Hipparcos} catalogue \citep{vanleuwen2007} and have reliable parallaxes. 65 co-moving stellar companions in a sample of 245 late-K to mid-M dwarfs yield a companion star fraction of $23.5\pm3.2$~per~cent within the sensitivity range.

The frequency of detected binaries with a given separation increases towards smaller separations (Fig.~\ref{fig:obsdist}). As measured by the Gaussian width, the fit to the \minms~separation distribution in \citet{wardduong2015} suggests that it is \emph{at least} $1.7\times$ wider than for the early \al~separation distribution, although the covered spectral-type ranges are comparable in both surveys. This potential tension will be qualitatively addressed in the discussion of our results (Sec.~\ref{sec:results}).

\section{A model for the Galactic field}
\label{sec:model}
\begin{figure}
 \includegraphics[width=0.45\textwidth]{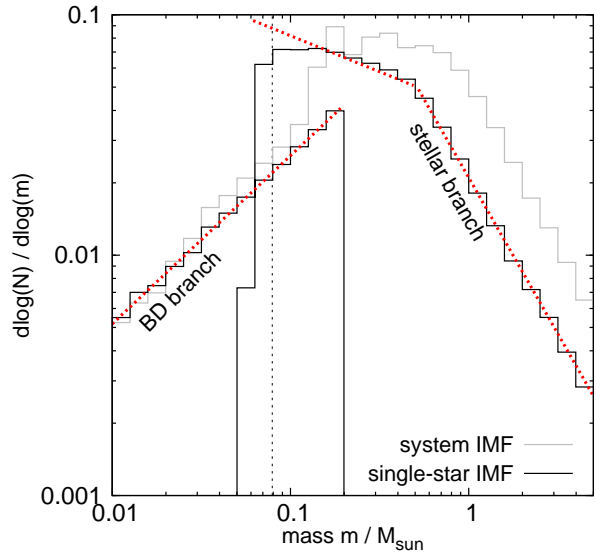}
 \caption{The stellar and substellar IMF are disjoint (black histograms). The dotted lines indicate an underlying power-law of the form $m^{-\alpha}$ with $\alpha=0.3,\;1.3$ and $2.3$ (from left to right). ``BD-like'' VLM stars are objects from the BD branch of the IMF lying above the hydrogen burning mass-limit (vertical dashed line) and ``star-like'' BDs stem from the stellar branch of the IMF lying below the hydrogen burning mass-limit. Both populations mix and form the system IMF ($31$\% and $\approx100$\% binary fraction for the BD and stellar branch, respectively, see Tab.~\ref{tab:params}). The discontinuity is thus hidden from an observer (grey histogram).}
 \label{fig:imf}
\end{figure}
The \al~M-dwarf \citep{janson2012m,janson2014m} and \minms~data \citep{wardduong2015} are for Galactic field binaries. A model is thus needed to describe how these emanate from their birth embedded stellar clusters. Such a DPS model is readily available from \citet[Section~\ref{sec:dynpopsynth}]{mk11}. A separate BD-like population (Section~\ref{sec:sepbdpop}) is here added to a star-like population.  For the star-like population, we assume universal initial conditions for late-type stars, and then account for the subsequent dynamical processing (Section~\ref{sec:process}).

The DPS model will provide \textit{predictions} for M-dwarf binaries in the Galactic field. These predictions do not stem from first principles because a theory of star formation does not exist which yields such information. Instead, the binary orbital-parameter distributions to start with before dynamical processing result from inverse DPS \citep{k95b} of the then available empirical data for solar-type stars \citep{DuqMay91} and pre-main sequence binaries \citep[e.g.][]{lz1993tau}. But since the involved DPS model parameters (Tab.~\ref{tab:params} below) have been constrained using independent samples, as explained in the forthcoming sections, the DPS model with a set of previously constrained parameters can be tested against other observations using the same parameters. This is done here for the \al~and \minms~data.

\subsection{Universal formation plus dynamical processing for late ``star-like'' objects}
\label{sec:process}
As a working hypothesis to address the question of the (non-)continuity of star formation the \textit{birth} binary population for star-like objects in embedded star clusters is assumed to be the same for all late-type stars \citep{k2011,kp2011}. ``Star-like'' refers to all bodies selected from the two-part IMF which extends into the BD regime (producing star-like BDs down to $0.06\msun$) and overlaps with a separate ``BD-like'' population \citep[see Fig.~\ref{fig:imf} and Section~\ref{sec:sepbdpop},][]{tk2007,tk2008tau}.

The birth binary population is built from an universal separation distribution for late-type stars derived by \citet{k95b} with an excess of long period (soft) binaries over that in the Galactic field \citep[eq. 8 in][]{k95b}, as seen in young star formation regions such as Taurus Auriga \citep[Tau;][]{kl1998,kraus2011tau} and Corona Australis \citep[CrA,][]{kn2008cra}. The initial binary population is defined using random pairing of binary component masses for low-mass stars ($\lesssim$a few $\msun$), which are selected from a standard two-part power-law stellar IMF \citep{k2001}, along with a thermal eccentricity distribution \citep[see also][]{k2013rev} which bends over in the course of dynamical processing \citep[their figs.~2 and~5]{mko11}. While random pairing does not well reproduce the rather flat mass-ratio distribution for solar-type stars and more massive primaries \citep{regmey2011,dk2013rev} it is in agreement with the weakly rising $q$-distribution for M-dwarf binaries and also with the overall mass ratio distribution of late-type stars \citep[see fig.~5 in][]{mk11}. The results on the semi-major axis distribution presented here are not affected by the chosen pairing mechanism since dynamical binary processing shows only a weak preference to disrupt low-$q$ systems first \citep{mko11,parker2013mr}. Any additional orbital parameters are calculated using Kepler's laws. The subsequent birth binary population is subjected to pre-mainsequence eigenevolution \citep{k95b} to account for gas-induced changes of orbital parameters in the circum-binary material during the cluster formation \citep[see also][]{stahler2010,kkp2012}. The \textit{initial} population, which is prone to dynamical processing inside its host environment, i.e. an embedded star cluster, is thus arrived at.

In order to calculate the effects of dynamical processing on initially binary-dominated populations inside young star clusters, we apply the analytical description of the dynamical processing in \textsc{Nbody} computations devised by \citet{mko11}. In terms of the resulting binary population, the outcome should be dynamically equivalent to an initially sub-structured and dynamically cold configuration, as \citet{pga2011} conclude. The initially binary-dominated population changes on a crossing-time scale, i.e the resulting distribution after a few Myr depends only on the initial stellar density \citep[see][]{mko11}.

\subsection{A separate ``BD-like'' population}
\label{sec:sepbdpop}
Given the existence of the BD desert and the apparently different binary characteristics between stars and BDs (Section~\ref{sec:intro}), \citet{tk2007,tk2008tau} quantified a stellar plus BD IMF which has a discontinuity around the hydrogen-burning mass limit. However, the stellar and BD parts have an overlap (Fig.~\ref{fig:imf}), since the star formation process does not care about hydrogen burning and it is unclear why the hydrogen-burning mass limit should constitute a sharp cutoff. Note that upon combining both populations an observer would not readily see the discontinuity but rather a declining and continuous IMF around the star--BD mass limit \citep[see Fig.~\ref{fig:imf} and][]{tk2007}.

In this contribution a ``BD-like'' population is added to the star-like population resulting from the Galactic field DPS model of \citet[][Section~\ref{sec:dynpopsynth}]{mk11}. A BD or late M-dwarf is called ``BD-like'' if it stems from the BD-like branch of the IMF (masses between $0.01$ and $0.2\msun$, Fig.~\ref{fig:imf}). The exact value of the upper cut-off mass for BD-likes is not known and might vary between regions. The best-fit values in \citet{tk2007,tk2008tau} for Tau, IC348 and the Trapezium cluster in the ONC are in the range $0.1-0.2\msun$, with individual uncertainties of about $\pm0.04$ to $\pm0.08$ (their table 3). For Tau its value might be closer to $0.1\msun$, and to $0.2\msun$ for IC348 and in the ONC, but their values agree within their $2\sigma$ mutual errorbars. Since Tau type aggregates cannot be dominant in contributing to the Galactic field, given a significant soft binary component not present in the field \citep{mk12,marks2014}, we here adhere to 
$m_{\rm max,BD}=0.2\msun$ determined for IC348 whose binary population resembles the Galactic field at long periods more closely \citep{mk12}. The most massive objects forming in the discs around host stars of $0.7\msun$ used in the analysis of \citet{li2015fragmentation} is truncated at $\approx0.2\msun$ as well, motivated by the mass spectrum of objects emerging in circumstellar discs in SPH computations of \citet{stamwhit2009discfrag}. The influence of the choice of the BD-like cut-off mass is discussed further in Sec.~\ref{sec:case}. Note that our chosen value must not be taken representative for any star forming region since we discuss here a BD population originating and superposed from many star forming events.

The BD-like population is excluded from {dynamical processing in its} birth cluster (as is described in Section~\ref{sec:process}), since we do not have good constraints on the initial BD population at this time. Note that this does \emph{not} imply that we consider this population to be dynamically inactive, i.e. we do \emph{not} assume the field population to resemble the primordial population. Instead the dynamical processing of the BD-like part is implicitely accounted for by constructing it such that upon superposing it with the dynamically processed star-like population, the BD binaries ($<0.08\msun$) added from both branches match the BD binary characteristics in the field. This is possible as the outcome of any dynamical model has to reproduce what is observed for BDs in the Galactic field as a constraint. This approach, however, fixes the parameters of the BD-like population and the dynamical population synthesis parameters a priori (Tab.~\ref{tab:params}). \emph{The here used BD-like population is 
thus the one for the Galactic field and must not be used for population synthesis in star clusters.} The BD-like population that should act as the input to star cluster models could be universal, but here we do not address this issue.

To model BD-like binaries, individual masses are selected from the BD IMF (Fig.~\ref{fig:imf}) randomly and stored in an array. From this array, masses are paired such that the observed mass-ratio distribution for BDs in the field, which has a strong preference toward unit mass-ratio, is reproduced. To do so, the biased-pairing algorithm introduced by \citet{thies2015} is used. This procedure applies a probability $p=q^{\gamma}$ to each binary, where $q=m_2/m_1\in[0:1]$ is the mass ratio and $\gamma=2$. If another random number is \emph{smaller} than $p$ the binary is accepted. If it is larger a new companion for the first selected object is assigned and the procedure is repeated. This ensures that a mass-ratio close to unity is strongly preferred for BDs. Note that a rejected companion is not discarded. It later becomes either a component of a different binary or stays single. This step is important to maintain the shape of the BD-like branch in the single-star IMF (Fig.~\ref{fig:imf}). Note also that biased pairing is a natural outcome of the formation of BDs in fragmenting stellar accretion discs \citep{thies2010bd}. Each BD-like binary is 
given a semi-major axis selected from the observed separation distribution of BDs in the Galactic field.\footnote{See Section~\ref{sec:case} for further discussion of this assumption.} Most of the selected BD-like objects remain single to yield the observationally constrained binary fraction of $\approx15$~per~cent for BDs upon combining it with the processed star-like population. The total binary fraction for BD-like binaries \emph{in the Galactic field} is then $\approx31$\%. The discontinuity in the IMF is measured by the fraction of BD-like to star-like objects in the population, ${\cal R}_{\rm pop}=N_{\rm BD-like}/N_{\rm star-like}$ and chosen to be $0.3$. This reflects the empirically determined value for Taurus-Auriga and the Pleiades \citep{tk2007}, i.e. for every three star-like bodies there is one BD-like object.

\subsection{Galactic field dynamical population synthesis}
\label{sec:dynpopsynth}
\begin{table}
\caption{Parameters defining the initial star-like population \emph{in embedded clusters}, the BD-like population \emph{in the Galactic field} and the Dynamical Population Synthesis procedure. From top to bottom the parameters denote (for details see the text): initial binary fraction and minimum mass for objects on the star-like branch, field binary fraction and maximum mass for objects from the BD-like branch, fraction of BD-like to star-like objects and biased pairing exponent, star formation rate in the Milky Way, power-law index of the ECMF, minimum embedded star cluster mass and typical embedded cluster half-mass radius.}
\label{tab:params}
\begin{center}
\begin{tabular}{c|c}
 Parameter & Value \\
 \hline
 \multicolumn{2}{c}{star-like population (embedded clusters)} \\
 \hline
 $f_b$ & $\approx100$\% \\
 $m_{\rm min}$ & $0.06\;\msun$ \\
 \hline
 \multicolumn{2}{c}{BD-like population (Galactic field)} \\
 \hline
 $f_b$ & $31$\% \\
 $m_{\rm max}$ & $0.2\;\msun$ \\
 ${\cal R}_{\rm pop}$ & $0.3$ \\
 $\gamma$ & $2$ \\
 \hline
 \multicolumn{2}{c}{Dynamical Population Synthesis} \\
 \hline
 SFR & $3\;\myr$ \\
 $\beta$ & $2.0$ \\
 $\meclmin$ & $5\;\msun$ \\
 $\rh$ & $0.1$~pc
\end{tabular}
\end{center}
\end{table}
{The stellar and BD population in a galaxy is the result of the addition of all populations formed in all embedded clusters.} According to the galactic field DPS model of \citet{mk11}, each embedded star cluster's binary population is processed for $3$~Myr, the time at which the first supernovae are expected to occur and in so doing drive out the residual-gas and destroy most of their natal stellar aggregate \citep{ll2003}.\footnote{The results are not very sensitive to this time-span as dynamical processing of initially binary dominated objects occurs rapidly on a crossing-time scale of the embedded cluster \citep{mko11}.} These clusters are thus the building blocks of the stellar single and binary population of the Galactic field. Embedded stellar cluster masses, $M_{ecl}$, are assumed to be distributed according to a single power-law initial embedded cluster mass function (ECMF), $\xi_{ecl}\propto M_{ecl}^{-\beta}$, with $\beta\approx2$ \citep{fuentemarcos2004,gieles2006}. The dynamically processed stellar population of a cluster with a given initial (embedded stellar) mass is weighted (multiplied) by the number of clusters that have this mass according to the ECMF. The sum of all these weighted populations results in the Galactic field population. As shown by \citet{mk11}, the DPS model works well to \emph{simultaneously} describe various orbital-parameter distributions of the Galactic field binary population \emph{with a single set of parameters}, depending on spectral-type (mass) of the primary component. This notion is supported by the present work and will be further substantiated in an upcoming contribution, which uses the latest observational data. In this work the DPS parameters inferred by \citet{mk11} will be used (Tab.~\ref{tab:params}).

\section{Results}
\label{sec:results}
\subsection{Separation distribution}
\label{sec:sepdist}
\begin{figure*}
 \includegraphics[width=0.45\textwidth]{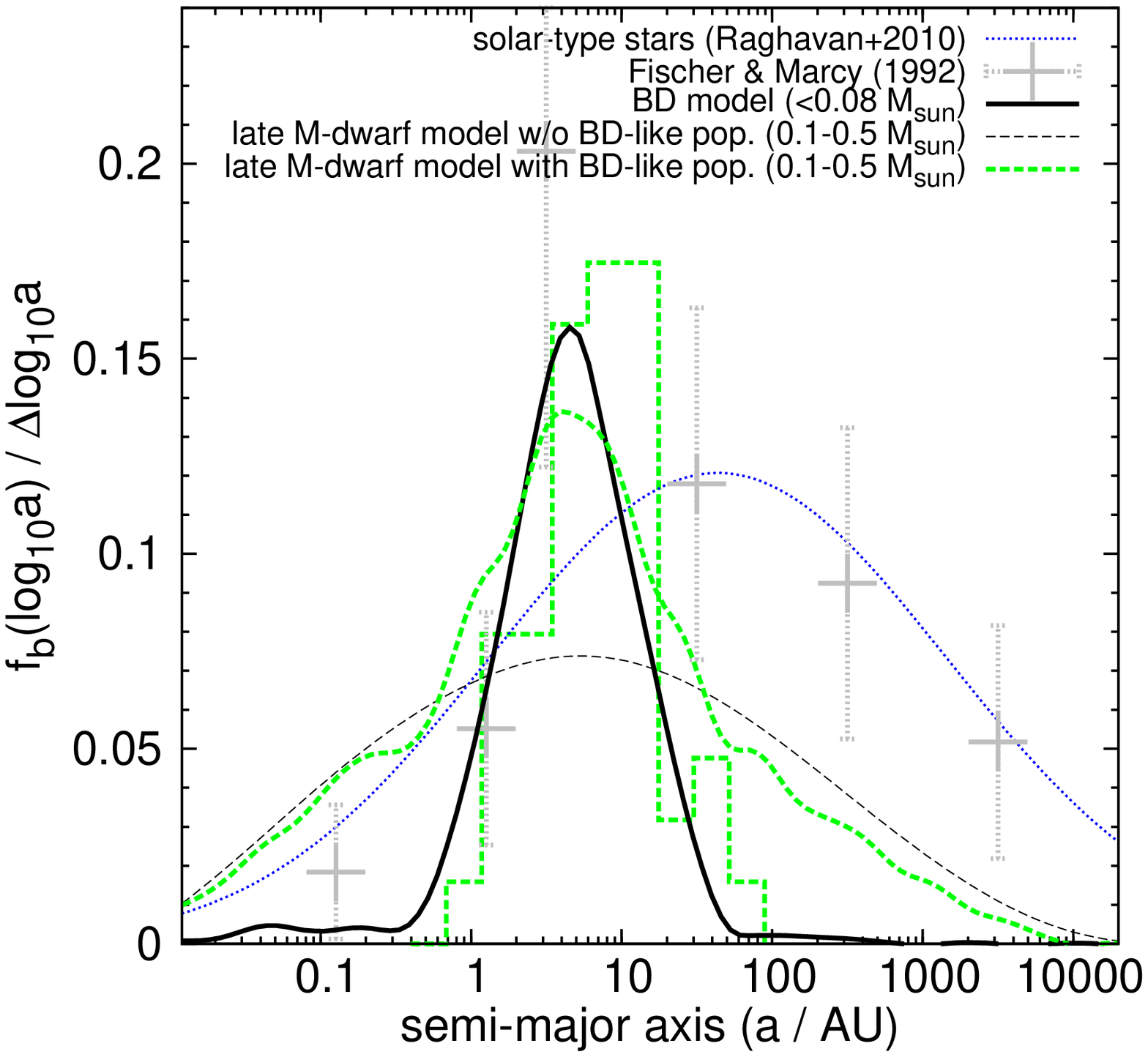}
 \includegraphics[width=0.45\textwidth]{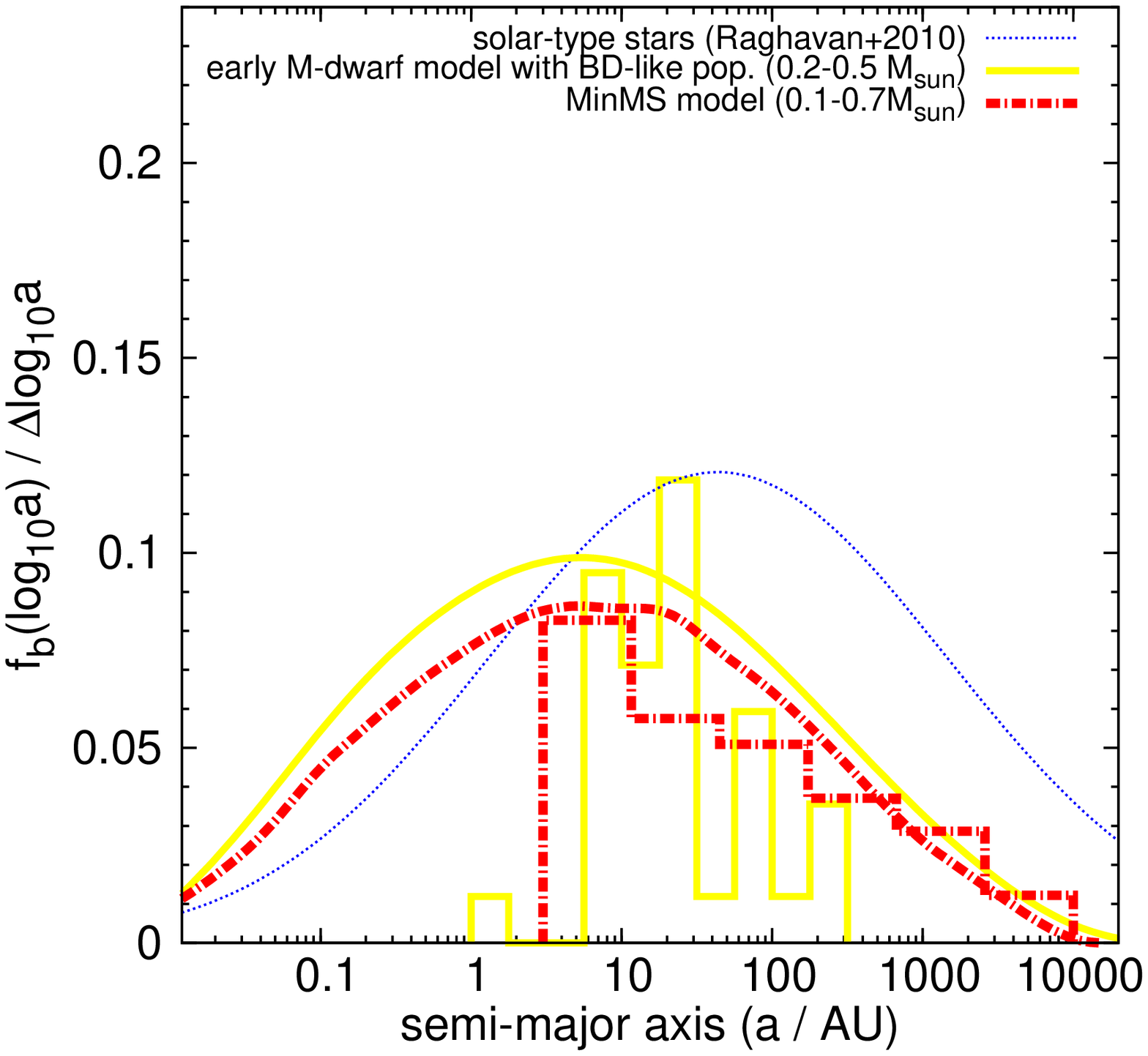}
 \caption{Comparing orbital semi-major axis distributions of the DPS model (curves) with \al~and \minms~data, respectively (histograms). For reference, the semi-major axis distribution of solar-type stars (dotted blue line) is shown in both panels. The DPS model binary mass ranges are chosen according to the mass estimates of the binaries in the \al~and \minms~survey. Line-types as in Fig.~\ref{fig:obsdist}. \textbf{Left panel:} Upon combining the star-like M-dwarfs (dashed black curve) with the BD-like M-dwarfs ($0.08-0.2\msun$ on the BD-like branch) in the DPS model, whose distribution resembles the one for (star-like + BD-like) model BDs (solid black curve), the late M-dwarf DPS model results (green dashed curve). It resembles the late-type data in the observed range (green dashed histogram, SC sample). For comparison, the M-dwarf data from the pioneering study by \citet[crosses]{fm1992m} is shown. \textbf{Right panel:} A peak is not expected (solid yellow and dashed-dotted red curve) in the early M-dwarf data (corresponding histograms) since the BD-like branch here extends to $0.2\msun$ only.}
 \label{fig:modeldatacomp}
\end{figure*}
Using a star-like formation mode only does not match the recent \al~data for late M-dwarfs. Fig.~\ref{fig:modeldatacomp} (left panel) however demonstrates that upon adding BD-like M-dwarfs, i.e. those late M-dwarfs which form as part of the BD branch in Fig.~\ref{fig:imf}, a peak on-top of the wider star-like distribution appears. This is due to, in the present formulation, the BD-like M-dwarfs sharing their separation distribution with those for BDs in the Galactic field. The DPS model binary fraction of $21.5$\% in the range $\log_{10}a=-0.4$ to $1.95$ is in reasonable agreement with the $17.9$\% observed in the late M-dwarf SC sample.

For a more sophisticated analysis, we compare the distribution shapes by means of a Kolmogorov-Smirnov (KS-)test. We do this by subjecting the DPS model to the completeness of the survey. One thousand random realizations of the DPS model are generated and the median of the match probability is adopted. This is the same method used by \citet{janson2014m} to find a Gaussian distribution that best describes the data. For details see their sec.~6.2. The test involves the DPS model mass-ratio distribution which is discussed in Section~\ref{sec:massratio}. The experiment results in a match probability of $30.4$\%, i.e. the hypothesis that the DPS model constitutes a parent distribution of the data is confirmed to within $\approx1\sigma$ certainty. This result is discussed further in Section~\ref{sec:discuss}.

It is interesting to note that the now more than 20-year old data of \citet{fm1992m} shows a similar excess of M-dwarf binaries which is located at about the peak of the field BD and late M-dwarf separation distribution. Primary masses in their observations were in the range $\approx0.1-0.6\msun$, i.e. a contribution of BD-like M-dwarfs is expected. The huge uncertainties did not suggest a real feature. But, retrospectively, as a tracer of a separate BD-like population, the peak might have been in front of our eyes all along.

For early M-dwarfs a similar peak on-top of a wider distribution is \emph{not} expected (Fig.~\ref{fig:modeldatacomp}, right panel). This is due to the BD-like branch extending, in the present formulation, up to $0.2\msun$ only (Fig.~\ref{fig:imf}). This value compares to the lowest-mass M-dwarfs in \citet{janson2014m}'s early sample. Note how the different height of the observed \al~early M-dwarf distribution is reasonably matched by the DPS model. The wings of the fitted distributions as seen in Fig.~\ref{fig:obsdist} are not reproduced, but with the exception of the lowest separation bin the DPS model might even compare with the raw \al~data. On the other hand, the available \minms~data and the distribution fitted to this data in Fig.~\ref{fig:obsdist} is in excellent agreement with the DPS model.\footnote{Note that, although \minms~targets with $<0.2\msun$ exist, we have run the DPS model without a BD population since only one binary with $<0.2\msun$ is part of the observed separation distribution.}

This immediately questions how the observational data on early M-dwarfs, which share similar properties (Sec.~\ref{sec:surveys}), relate to one another. While the increased frequency of binaries with distant companions can be readily explained through the enhanced coverage of semi-major axes in the \minms~sample, the conflict remains for the closest covered separations (Fig.~\ref{fig:obsdist}). This is at the very least puzzling since the covered spectral-type range is comparable in both surveys. If the slightly wider mass-range of the \minms~targets is responsible, yet unknown processes could be at work around the M-/K-dwarf boundary. Another possible caveat is the systematic age difference between the samples (Sec.~\ref{sec:al}). However, at present it is hard to see a reason how and why these differences should approximately double the observed distribution width.

Alternatively, one may ask whether the fits of Gaussian distributions to the \al~and \minms~data are reliable representations of the underlying parent distributions. The raw separation distributions (histograms in Fig.~\ref{fig:modeldatacomp}, right panel) appear to compare better in the separation range where both studys overlap than a comparison of the fitted distributions suggests. Is it then possible that both distributions are not distinct after all and stem from a common parent distribution? In this latter case the Gaussian fits would be too simplistic. We cannot decide this on the basis of the present study and possibly requires further observations. It is noted that neither \citet{janson2012m} nor \citet{wardduong2015} attempt to make any detailed fit involving thorough statistical tests but merely discuss various general options. So it is as of yet not clear whether a parent distribution exists that fits both datasets simultaneously at acceptable confidence limits. A wider distribution for early M-dwarfs as in the \minms~data compares better with some previously obtained results \citep{fm1992m,d2004m}, while a narrower distribution as in the early \al~data is suggested as well by \citet{bergfors2010m}'s study. Note that the latter two studies which find a narrow distribution used the same observational technique, i.e. lucky-imaging.

\subsection{Mass ratio distribution}
\label{sec:massratio}
\begin{figure}
 \includegraphics[width=0.45\textwidth]{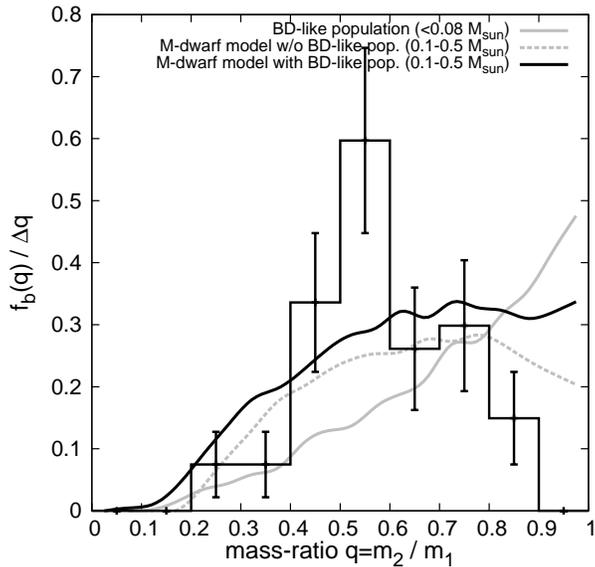}
 \caption{Oberserved and DPS model mass-ratio distributions for binaries in the semi-major axis range $\log_{10}(a/AU)=-0.4$ to $1.95$, as in the observations. Combining star-like (dashed grey line) and BD-like M-dwarfs (solid grey line) as in Fig.~\ref{fig:modeldatacomp}, the late M-dwarf DPS model appears (solid black curve). The histogram with Poissonian errorbars shows the late \al~data (SC sample). Caution is required since mass estimates for \al~targets are uncertain (Sec.~\ref{sec:massratio}).}
 \label{fig:massratio}
\end{figure}
Masses and mass ratios are difficult to estimate for the VLM objects in the \al~survey. The dominant difficulty stems from the fact that the ages of the stars are only very loosely constrained, which means that the transition from a star's brightness to its mass will be accompanied by very large uncertainties. Furthermore, the evolutionary and atmospheric models \citep{hauschildt1999,baraffe2003,allard2014} that are used to make such a transition are themselves uncertain, since they have not been calibrated against observations for large sections of the parameter space. Both of these issues are particularly critical for masses that approach the BD range. This is why the observed mass-ratio distribution should not be too much relied on. Regardless, as already stated, the observed features of interest for our purposes are present in the data at a statistically significant level.

In the DPS model, mass ratios for BD-like objects prefer values close to unity, as observed for BDs in the Galactic field. The DPS model facilitates this through biased-pairing (Section~\ref{sec:sepbdpop}). Fig.~\ref{fig:massratio} demonstrates how combining this with star-like M-dwarfs leads to a distribution that increases with increasing $q$ and flattens beyond $q\approx0.5$, for binaries in the semi-major axis range $\log_{10}(a/AU)=-0.4$ to $1.95$.

\subsection{A BD desert in the data?}
\label{sec:desert}
\begin{figure}
 \includegraphics[width=0.45\textwidth]{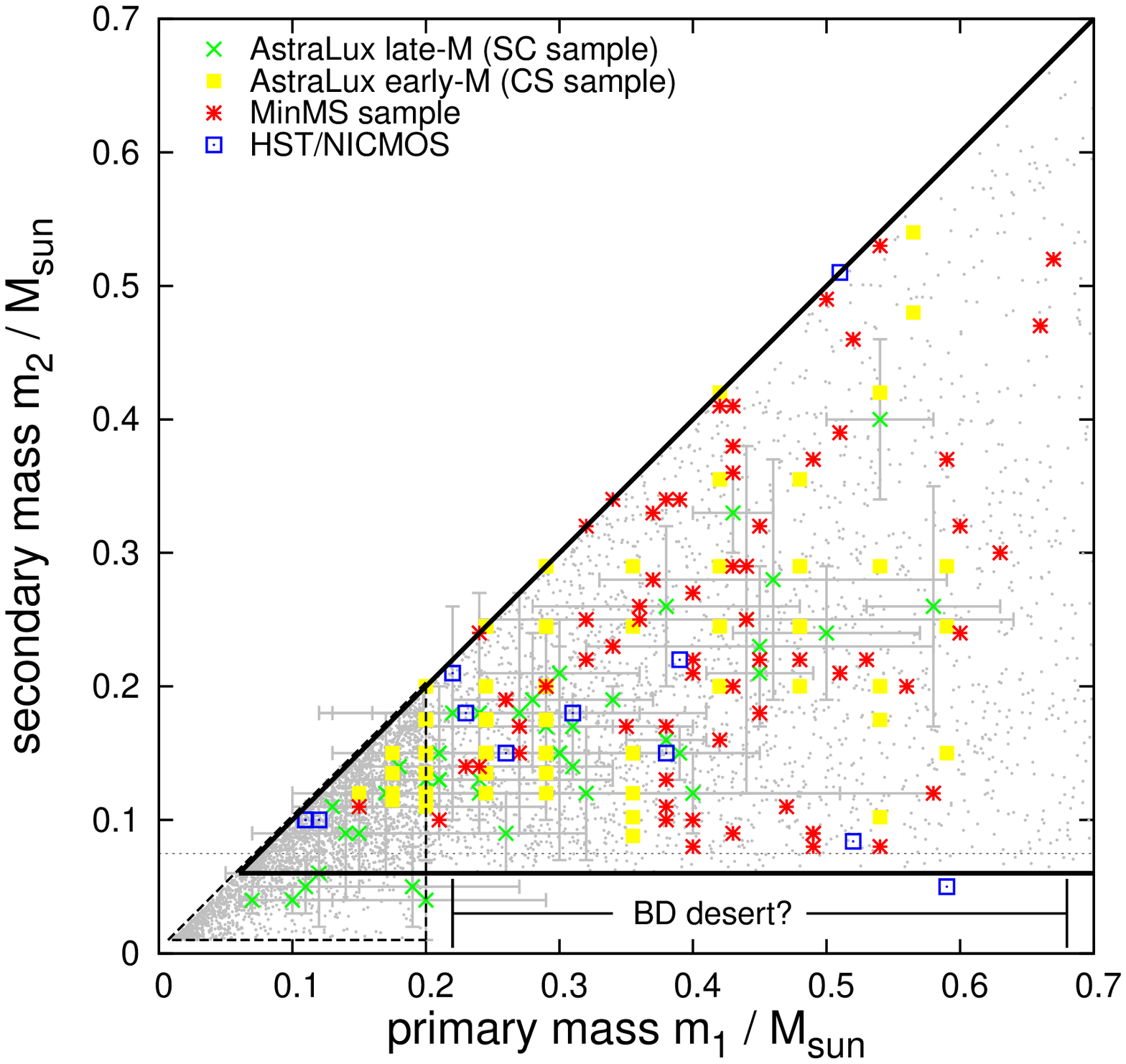}
 \caption{Primary vs. secondary component masses in the \al~\citep{janson2012m,janson2014m} and \minms~M-dwarf data \citep{wardduong2015}, as well as the HST/NICMOS VLM star data \citep[squares,][]{dieterich2012}. Color-coding as in Figs.~\ref{fig:obsdist} and~\ref{fig:modeldatacomp}. The DPS model prediction is depicted as grey dots. The density of dots is a measure of the expected numbers. VLM M-dwarfs and BD binaries have a tendency toward unit mass-ratio. The upper solid triangle contains binaries from the star-like branch, and the lower dashed triangle contains binaries from the BD-like branch of the IMF. Where the triangles overlap, binaries from both branches can be found. The horizontal dotted line is the hydrogen burning mass-limit which roughly marks the companion detection limit for the early M-dwarf surveys.}
 \label{fig:desert}
\end{figure}
If BDs form in the same way as stars the natural expectation is that BDs pair in the same way with stars as stars do among each other. And if the IMF is continuous over the hydrogen-burning mass limit BDs should be the most abundant companions to stars \citep{kb2003tau}. However, studies of VLM objects have demonstrated that this is not the case, i.e. a BD desert is apparent \citep{mzb2003desert,grether2006desert,dieterich2012}. The SPH plus \textsc{Nbody} approach of \citet{li2015fragmentation} to study the BD-like formation mode, i.e. disc fragmentation, is in agreement with the observed BD desert. Although \citet{grether2006desert} suggest that a BD desert is a natural consequence of a universal companion mass function (CMF), \citet{dieterich2012} show, using their \textsc{HST/NICMOS} sample, that VLM stars have a tendency toward unit mass ratio, i.e. BD companions are rare if not absent. This suggests that deviations from a universal CMF likely exist, at least for the BD regime (see their \S6.3 and \S6.4 for a discussion).

Being based on random pairing for star-like objects and biased-pairing for BD-likes, the DPS model suggests a similar BD desert to be present. Fig.~\ref{fig:desert} shows that the DPS model prediction is overall consistent with the \al~and \minms~data. The more the primary component mass approaches the hydrogen-burning limit from above, the closer the mass-ratio gets to unity. This is consistent with the finding of \citet{dieterich2012}. As the primary mass increases, the DPS model suggests that systems with low mass ratios should exist, but they are not seen in the late \al~observations. This might reflect the difficulty to detect systems where the brightness of the primary outshines a potentially present lower mass star or BD. \emph{Thus, the DPS model predicts that as instrument sensitivity increases, the observed low mass-ratio systems should become apparent for late M-dwarfs, increasing the observed binary fraction and bringing it closer to that predicted by the DPS model.} Such systems have been found in the \textsc{HST/NICMOS} sample \citep[squares in Fig.~\ref{fig:desert}]{dieterich2012}.

Whether a similar BD desert exists for this late M-dwarf sample is thus difficult to assess with the \al~data. If a desert were to exist, it would contradict the suggestion that the semi-major axis distribution constructed from the very same data (Figs.~\ref{fig:obsdist} and~\ref{fig:modeldatacomp}) is a tracer of a common, canonical formation mode for all stars and BDs. Detailed observational investigations of the BD desert for VLM stars \emph{in conjunction} with their separation distributions will thus provide important insights into the issue of the (non-)continuity of star formation.

\subsection{Complementary M-dwarf studies}
\begin{figure}
 \includegraphics[width=0.45\textwidth]{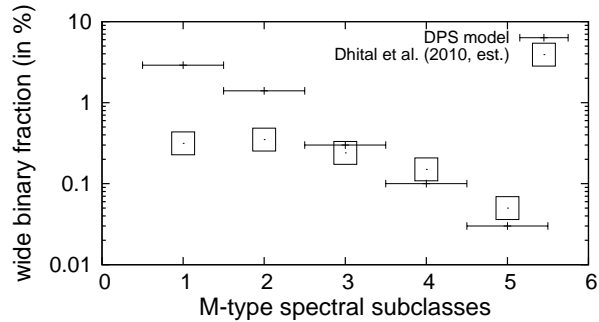}
 \caption{DPS model binary fractions and those observed for wide binaries in the study of \citet[estimated from their fig.~16]{dhital2010widemdwarfs}. The semi-major axis range covers $\log_{10}(a/AU)=3$ to $5$. Beyond spectral type M5 the binary fraction is essentially zero, both in the study and in the DPS model. Observed binary fractions are lower limits.}
 \label{fig:wide}
\end{figure}
The \textsc{SLoWPoKES} survey \citep{dhital2010widemdwarfs} investigates binarity of low-mass, wide common-proper motion binaries in a catalog from the Sloan Digital Sky Survey (SDSS). In Fig.~\ref{fig:wide} we compare their observationally deduced wide binary fractions, $a=10^3-10^5$~AU, with the DPS model. A similar declining trend towards later spectral types is seen for both the DPS model and the observations. In the DPS model context this is due to the easier break-up of binaries with lower mass primaries in their birth clusters since their binding energy decreases and the uneven mass-ranges covered in each spectral subclass \citep[shrinking towards later M-dwarfs, e.g.][]{baraffe1996specclass}. The DPS model appears to overestimate the published binary fraction for the earliest M-dwarfs by about an order of magnitude. \citet{dhital2010widemdwarfs} state that their wide binary fraction is likely a lower limit because observational biases and incompleteness play a significant role. The excess of model binaries might be related to this. If not, then this might indicate a short-coming of the DPS model. We note, however, that for M3 and later spectral types, the DPS model and observed fraction agree better.

At the small semi-major axis end the spectroscopic study by \citet{clark2012closemdwarfs} finds a binary fraction of $2.9^{+0.6}_{-0.8}$~per~cent for cool M-dwarfs binaries from the SDSS having $\log_{10}(a/AU)<-0.4$ \textit{assuming} a uniform prior semi-major axis distribution \citep[unlike the rising semi-major axis distribution of][]{k95b}. This value is in agreement with the DPS model fraction of $2.4$~per~cent in the range $-1<\log_{10}(a/AU)<-0.4$. \citet{clark2012closemdwarfs} additionally show that the spectroscopic binary fraction is a continuously increasing function of primary mass, from BDs to massive O-type stars, which is an additional test of the DPS model. Since the present work is on M-dwarfs, we investigate this dependence in a forthcoming contribution.

\section{Discussion}
\label{sec:discuss}

\subsection{A case for separate populations}
\label{sec:case}
The sole fact that the most straight-forward strategy to combine star-like and BD-like objects is able to reproduce the observations is remarkable. If one had thought of the implications of a separate BD-like population on the late M-dwarf separation distribution \emph{before} the \al~data became available, the DPS model described in Section~\ref{sec:model} using the parameters in Tab.~\ref{tab:params} would have been devised in exactly the same way. \emph{No attempt at trying to merely reproduce the observations has been made here.} Having said this, the DPS model thus post-hoc \emph{predicts} both the \al~late M-dwarf binary fraction and semi-major axis distribution, the possibly different distribution amplitudes between the early and late M-dwarf data, and the \minms~data for early M-dwarfs.

The match probability of $\approx30$\% is already good. It might increase by relaxing the \emph{assumption} that the whole BD-like population, which includes late M-dwarfs from the BD-like branch, follows the BD separation distribution in the Galactic field. A proper initial distribution for BD-like objects in embedded star clusters which has evolved alongside the star-like objects through dynamical processing is not yet available. But since VLM M-dwarf binaries from the BD-like branch are on average more strongly bound than binaries with a BD primary, it is to be expected that in reality M-dwarf binaries from the BD branch with a larger separation can better survive dynamical processing in embedded clusters. Their separation distribution in the Galactic field will thus likely extend to somewhat larger semi-major axes than assumed here, and the amplitude of the peak will increase as well. This would, in turn, improve the consistency between the DPS model and the observations.

The DPS model requires an overlap of the star-like and BD-like branches of the IMF (Fig.~\ref{fig:imf}). For the present modelling the BD-like branch is assumed to extend to $0.2\msun$, as empirically determined for IC~348 (Sec.~\ref{sec:sepbdpop}), which is consistent with the mass spectrum arising for objects forming through disc fragmentation in the SPH computations of \citet{stamwhit2009discfrag}. If the cut-off mass for the BD-like branch is lowered to, say, $\approx0.12\msun$ (the best-fit value for Tau) the amplitude of the peak in Fig.~\ref{fig:modeldatacomp} is lowered as well. This is due to the smaller overlap of the BD-like branch with the observed range of primary masses in the late \al~data which extends down to $\approx0.1\msun$ only. The match probability then changes to $11.6$~per~cent, thus discarding the DPS model is possible at less than $2\sigma$ confidence. Staying within the DPS model context, the lower match probability using a lower cut-off mass as determined for Tau might simply imply that Tau-like binary formation is not dominant in contributing to the field, as is additionally evidenced by its super-field binary 
fraction for long-period binaries \citep[see Sec.~\ref{sec:sepbdpop}]{kl1998,kraus2011tau}. Though this cut-off mass is probably not significantly larger, we note that the uncertainty associated with this empirically-determined parameter implies that a somewhat larger value is also consistent with the data (\ref{sec:sepbdpop}), and this could improve the agreement with the DPS model.

Uncertainties in the binary fraction of Galactic field BDs and in the peak and width of the BD separation distribution are not considered here. In principle, with the available observations we have the freedom to vary these parameters within the uncertainties to further improve agreement with the data. However, we refrain here from arbitrarily varying these parameters to obtain a better match since the reasonable agreement with this simplest DPS model speaks for itself. Instead the issue will be revisited once constraints for an initial BD-like population become available.

\subsection{On the meaning of ``Universality''}
\label{sec:meaning}
The notion of universal binary formation is that star formation leads to invariant formal distribution functions due to physical processes like energy and angular momentum conservation and the chemistry of molecular clouds, all of which are the same everywhere, except perhaps in very intense star bursts. This implies an \emph{environment-independence} of binary formation (Sec.~\ref{sec:environment}), potentially similar to the idea that universal star formation leads to an invariant formal distribution of stellar masses, the IMF. These distribution functions are parent distribution functions, from which a particular case is discretised, or rendered. Universal binary formation does not preclude, however, a dependence on the primary component mass, e.g. different birth binary distributions for BDs and stars and in different mass ranges ($0.06-1.5\msun$, $1.5-10\msun$, $>10\msun$) are possible (Sec.~\ref{sec:massindepence}). The universal distribution functions can be \emph{primary mass-dependent} by, e.g., adding further formation channels, like circumstellar disc fragmentation for BDs, which do not change the underlying universal physical processes. We need such functions to initialise, for example, \textsc{Nbody} models in order to study how young and old clusters evolve into the field and associations.

\subsubsection{Environment (in-)dependence?}
\label{sec:environment}
An environment-independence of binary formation has often been suggested but the observational data don't appear to be conclusive \citep{k2011,kp2011,mk12,King2012b,marks2014,parker2014,leigh2014bingc}.

\citet{King2012b} found the orbital-separation distributions in seven young star formation regions to be statistically indistinguishable, consisting mostly of hard binaries given the regions' presently observed conditions. By implicitly assuming that the regions were not denser in the past they concluded that the observed distributions resemble the ones at birth. On the other hand, \citet{mk12} demonstrated that these same regions are consistent with dynamically processed universal birth distributions for late-type stars \citep{k95b} if they were significantly denser in the past. \citet{marks2014} later re-visited this issue, and argued that the two competing scenarios are statistically indistinguishable, given the low number of observed binaries in these regions. It follows that solutions to the observations are degenerate.  In order to settle the issue of whether or not star formation in these regions was the same with large initial binary fractions, we need to know if these regions were significantly 
denser in the past. With this in mind, \citet{marks2014} offered several arguments to help constrain the initial cluster densities, and concluded that such a scenario is indeed plausible.

\citet{parker2014} offered an intriguing suggestion to break this density degeneracy. It involves measuring a star forming region's degree of substructure, and asking if it can result from dense initial conditions. The author assumes the \citet{k95b} primary-mass independent initial separation distribution for late-type stars that is evolved in dense, initially substructured and subvirial/collapsing clusters using \textsc{Nbody} models. The simulations produce clusters that are too centrally concentrated, compared to actual observations of star forming regions (except maybe the Orion Nebula Cluster, ONC). \citet{parker2014} finds initial density estimates from a comparison to the region's observed structure that are smaller than those constrained by \citet{mk12} from a comparison to the region's observed binary separation distributions. Unfortunately, the author does not compare the resulting separation distribution in his intermediate density computations -- which reasonably match the present-day structure -- to the observed distributions. Interestingly, fig.~1 in \citet{parker2014} suggests that these computations produce separation distributions that lie between those observed in Tau or CrA (i.e. an excess of long period binaries compared to the field, as in the \citet{k95b} distribution) and the ONC (i.e. depleted in long-period binaries). This is where the observed separation distributions in the remaining investigated regions lie as well \citep{mk12,marks2014}. If the binary populations in the intermediate density computations of \citet{parker2014} were indeed to resemble the observed ones, as we here suggest, they seem to support the hypothesis of environment-independent birth distributions for late-type stars. This is because the computations then simultaneously reproduce the observed present-day structure and binary populations. This raises the question: Why do \citet{mk12} and \citet{parker2014} find different initial cluster densities? One possible contributing factor is the different initial cluster setups (spherical and virialized vs. substructured and collapsing, respectively). However, \citet{pga2011} conclude that this should not make any difference as far as the dynamical processing of the binary population is concerned.

\citet{leigh2014bingc} confirm a similar density degeneracy of the primordial binary population for globular clusters (GCs). Using MOCCA computations \citep{giersz2013mocca} over a Hubble time they show that solutions to reproducing simultaneously the rather low binary fractions and an anti-correlation between the binary fractions and masses of GCs \emph{inside} the half-mass radius are degenerate in terms of the initial GC densities and initial binary fractions, quite similar to the density degeneracy for young star forming regions. However, they break this degeneracy by demonstrating that only dense initial configurations, which match densities observed for potential GC progenitors, in combination with large initial binary fractions account for a similar anti-correlation seen \emph{outside} the half-mass radius in Galactic GCs. Thus, the observations for GCs are consistent with the universality hypothesis \citep[not precluding different origins,][]{leigh2012,leigh2013,leigh2014bingc}.

\subsubsection{Mass-independent or field-like binary formation?}
\label{sec:massindepence}
The present work suggests that primary-mass independent birth distribution functions for late-type binaries plus dynamical processing continues to be a valid working hypothesis for M-dwarf binaries. Whether the birth distribution functions can be primary-mass dependent for late-type binaries or not is being debated.

In a recent contribution, \citet{pm2014} find that their \textsc{Nbody} computation with fractal initial conditions and peak densities around $1000\mpc$ do not reproduce observations of binaries in the Galactic field, when using the \citet{k95b} distribution as input and \emph{assuming BDs as a continuation of stars with the same universal initial conditions}. This has been done before in \citet{kroupa2003bd}, who show that this leads to distributions inconsistent with observations, and in particular too many star-BD binaries. \citet{kroupa2003bd} concluded that BDs need to be treated with their own pairing rules.

More precisely, in the computations of \citet{pm2014}, the orbital separation distributions after $10$~Myr of dynamical processing do not match the Galactic field separation distributions for
\begin{enumerate}
 \item brown dwarfs \citep{tk2007},
 \item M-dwarf binaries in the \al~survey \citep{janson2012m,janson2014m},
 \item solar-type binaries \citep{r2010} and
 \item A-type binaries in the VAST survey \citep{derosa2014vast}.
\end{enumerate}
Based on this, \citet{pm2014} suggest that the initial distributions are primary-mass dependent. Thus, computations with the observed distributions as input to the otherwise same cluster provided a better match to the observations, although not perfectly due to the dynamical break-up of wide binaries among the G- and A-type population. They conclude that the field binary populations are indicative of the star formation process in clusters.

\begin{figure}
 \includegraphics[width=0.45\textwidth]{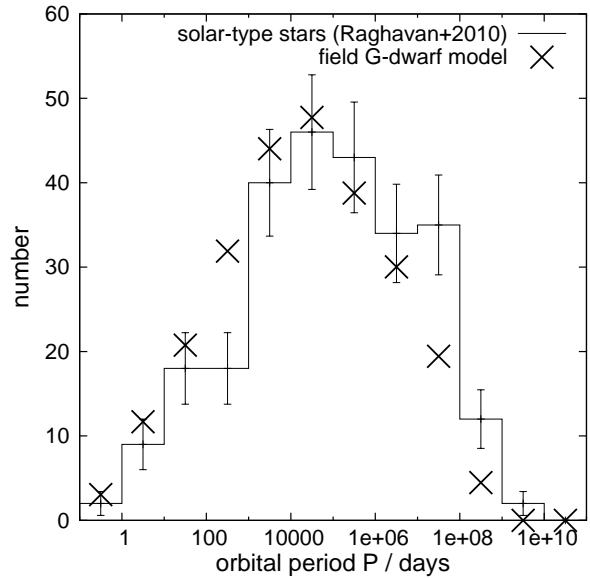}
 \caption{Comparison of orbital period distributions for solar-type stars \citep[histogram with Poisson errors,][]{r2010} and G-dwarfs in the DPS model (thick crosses). The same parameters (Tab.~\ref{tab:params}) used to construct the M-dwarf DPS model (Fig.~\ref{fig:modeldatacomp}) are employed.}
 \label{fig:solar}
\end{figure}
The results presented in this paper are consistent with a different interpretation, however, which have the advantage of being \emph{predictive}. In the following we address the apparent disagreement with the four populations above:
\begin{enumerate}
 \item a meaningful comparison of the dynamically processed universal semi-major axis distribution for late-type stars and the BD binary population in the Galactic field relies on the assumption that star formation is continuous and that the \citet{k95b} distribution is valid for BDs \citep{pm2014}. If this were not true, the comparison would not be meaningful. As demonstrated here, requiring a separate BD population as in \citet{kroupa2003bd} and \citet{tk2007,tk2008tau} yields agreement with the observed BD binary semi-major axis distribution by construction. Importantly, the \citet{k95b} distribution was not developed to match the observed properties of binary BDs, but was instead developed for (originally only) solar-type stars after dynamical processing.
 \item as shown in this contribution, the universality hypothesis for late-type stars following the \citet{k95b} orbital parameter distributions, combined with dynamical processing and a separate BD-like population, can successfully reproduce the late \al~and early \minms~M-dwarf data.
 \item extracting the orbital period distribution for G-dwarfs in the Galactic field from the DPS model adopting the same parameters used to extract the M-dwarfs (Tab.~\ref{tab:params}) shows agreement between the observed period distribution of solar-type binaries in the Galactic field \citep[including orbits in hierarchical multiples]{r2010} and the DPS model (Fig.~\ref{fig:solar}). This is not surprising, though, since the \citet{k95b} model was designed to match the solar-type data of \citet{DuqMay91}, and their separation distribution of orbits is indistinguishable from the one obtained by \citet{r2010}. On the origin of the different conclusions reached by \citet{pm2014} we can only speculate. We note that \citet{k95b} explicitly demonstrated that only certain combinations of initial cluster masses and radii reproduce the observed pre-mainsequence and field binary populations simultaneously. This yields constraints on the physical conditions in correlated star-forming events (i.e. embedded clusters).
 \item while A-type binaries in the \textsc{VAST} survey show a narrow distribution peaked around $370$~au \citep{derosa2014vast}, spectroscopic A-type binaries are also known to exist at shorter separations \citep{abt1965a,carrier2002a,cp2007a}. Even a double-peaked distribution for A-type binaries is suggested \citep{dk2013rev}. However, spectroscopic data is needed to complement \citet{derosa2014vast}'s data for a complete picture. We will address the \textsc{VAST} data in future work. However, the (initial) distributions for O and B-type binaries appear to be different from that given in \citet{k95b}. This may be indicative of a different formation channel for O- and B-type stars, e.g. through the star-formation process at the density centres of proto-clusters \citep[e.g.][]{bb2005}, analogous in some ways to a separate BD formation channel. In this case, the DPS model should be extended to account for observations of binaries with high-mass primaries \citep[e.g.][]{sana2012o}, as developed by \citet{okp2015}. However, one would naively expect any change in the star formation process to be mapped onto the stellar IMF \citep[cf.][]{bk2012r136}.
\end{enumerate}

The take-away message is that both a field-like formation scenario (``what-you-see-is-what-you-get'') and the DPS model with a separate BD-like population explain the Galactic field observations. The latter formulation, assuming one birth binary population distribution function for low-mass stars ($\lesssim1.5\msun$) plus dynamical processing to be the origin of all late-type stellar binaries and adding a separate BD-like population (and perhaps a separate OB population) explains the data. This scenario is not only consistent with, but actually explains the large range of estimated global binary fractions, from $\lesssim10$~per~cent in $\omega$~Centauri \citep{e1995wcen} to sub-field fractions in other GCs \citep{m2012acs} to $\gtrsim90$~per~cent in young star forming regions \citep{kl1998,d1999}. Observations of separation distributions in the latter objects might support this notion as well \citep{kp2011,mk12,marks2014}. In addition, very young regions like Tau and CrA \citep{kl1998,kn2008cra,kraus2011tau} as well as 
proto-stellar binaries \citep{connelley2008proto} exhibit a binary excess at long periods that do not support a field-like formation scenario. These populations all have a significant soft component, which changes quickly through dynamical processing in intermediate to dense environments. Such observations result naturally from different degrees of dynamical processing of the underlying binary populations.

Thus, the dynamical modification of the \citet{k95b} distribution combined with a separate BD-like population is not only consistent with the data, if the initial cluster density is high, but it also allows binary populations to be predicted in the Milky Way and other galaxies \citep{mk11}. The binary excess in the youngest and sparsest star forming regions is accounted for by the \citet{k95b} distribution used in the DPS model, and its dynamical modification yields the binary populations in star clusters and star forming regions \citep{mko11,mk12,marks2014,leigh2014bingc} and in galaxies \citep[and this contribution]{mk11}.

We caution that taking observed distributions and using them as input for computations of clusters which resemble the presently observed state as initial conditions (or even a moderately\footnote{I.e. a configuration which does not place the hard-soft boundary below about the maximum of the input semi-major axis distribution.} dense configuration) will in most cases not change the observed binary distributions significantly, unless the observed cluster is extremely young ($\ll1$Myr) such that the binary population did not have the time yet to adapt to its host cluster properly. This is because dynamical binary processing occurs rapidly if the population had been initially binary dominated and will rather quickly find an equilibrium with its host environment \citep[the changes happen on a crossing-time scale,][]{mko11} before the long-term, two-body relaxation driven evolution dominates further processing \citep{leigh2014bingc}. Thus, such an equilibrium state is what one will most likely observe.

\subsection{Predictions}
Whether BDs and stars are indeed separate populations with their own initial mass and binary distribution functions can be constrained with future surveys of Galactic field M-dwarf binaries. The presence of a separate BD-like subpopulation unmasks itself through a peak located close to the field BD separation distribution which resides on-top of a wider distribution, as a result of the superposition of the star-like and the BD-like M-dwarfs (Fig.~\ref{fig:modeldatacomp}, left panel). The DPS model predicts that M-dwarf binaries in the field exist in significant numbers outside the sensitivity range of the \al~survey \citep[e.g. as in][]{fm1992m,d2004m}, i.e. the separation distribution is wider than reported by the \al~data, both for early \citep{janson2012m} and late \citep{janson2014m} M-dwarfs. The recently published \minms~data \citep{wardduong2015} supports such a notion for the early M-dwarfs. If true, the late M-dwarf data roughly reproduce the peak of the DPS model distribution, whereas the observed early M-dwarf data missed it slightly due to the sensitivity constraints.

Here the BD-like branch of the IMF extends up to $0.2\msun$. All M-dwarf binaries with $\lesssim0.2\msun$ in the $\approx1-10$~AU range are thus apriori BD-like binary candidates. By comparing these objects among each other, observers might be able to find further signatures for a BD-like formation mode, aside from the peak in the separation distribution. These signatures might be hidden, e.g., in other orbital parameters such as their eccentricity. If the true BD-like M-dwarf binaries follow the Galactic field BD mass-ratio distribution they preferentially have a mass-ratio
close to unity. The analysis of disc fragmentation in \citet{li2015fragmentation} provides further clues.

We expect a peak superposed on a wider distribution in M-dwarf binary populations which have primary components with masses below $0.2\msun$. The peak is not expected if such binaries are absent, as is e.g. the case for \citet{janson2012m}'s early M-dwarfs and the \minms~sample \citep{wardduong2015}. Their sensitivity limits do not include the separation range where the peak is expected and prevents us from testing this scenario. Should a peak be seen for early M-dwarfs in later surveys as well, the maximum mass of BD-like objects needs to be shifted to larger masses. \emph{This provides thus a means to constrain observationally how deep the separate BD-like population, if it exists, penetrates into the star-like regime.}

Our random pairing scenario of binary components for star-like M-dwarfs predicts that low mass-ratio late M-dwarfs will be detected at larger primary masses and that these will be lacking BD companions (Fig.~\ref{fig:desert}), i.e. extend the BD desert. These are currently invisible due to \al~instrumental sensitivity constraints.

\section{Summary}
\label{sec:sum}
It has been demonstrated how combining a dynamically processed, initially binary-dominated universal star-like population for late-type stars ($\lesssim1.5\msun$) with a separate BD-like population forming through circumstellar disc fragmentation, which produces some VLM stars, leads to a prediction for the semi-major axis distribution of late M-dwarfs in the Galactic field. Upon adding the two formation channels, Dynamical Population Synthesis \citep{mk11} produces a narrowly peaked distribution that appears on-top of a wider distribution. The simplest DPS model resembles the late M-dwarf data obtained in the \al~survey \citep{janson2014m} and the null hypothesis that the DPS model provides a parent distribution for the observations is confirmed to within $1\sigma$ certainty. It has been discussed that agreement between DPS model and observation could improve once an estimate for an initial BD-like binary population becomes available and once it is allowed to participate in dynamical processing alongside the star-like objects in the DPS model. As a byproduct our results demonstrate that the hypothesis of primary-mass independent period, or semi-major axis distribution functions accounts for all late-type populations in the Galactic field. The opposite conclusion reached by \citet{pm2014} does thus not follow unambigiously from the data (see Section~\ref{sec:massindepence} for a discussion).

The DPS model predicts that binaries outside the sensitivity range of the \al~survey will be detected. A peaked semi-major axis distribution superposed on a wider distribution as for the late M-dwarfs is \emph{not} expected for early M-dwarfs \citep[as those in][]{janson2012m,wardduong2015}. Once surveys for early M-dwarfs become sensitive to smaller separations than covered by the \al~and \minms~data this notion will provide a means to measure how deeply the BD-like branch penetrates into the stellar regime. The DPS model separation distribution agrees excellently with the early M-dwarfs in the \minms~sample but less well with the early \al~data. It is pointed out that the observed separation distributions for early M-dwarfs found by \citet{janson2012m} and \citet{wardduong2015}, respectively, are potentially in tension (Sec.~\ref{sec:sepdist}).

Given the surveys of nearby stars, a BD desert is expected to be present in the \al~data. No BD companion is seen in the data above a primary mass of $0.2\msun$ but the decreasing detection probability of a low mass-ratio system with increasing primary mass does not yet allow for any firm conclusions. If a BD desert in the data were confirmed it would contradict the continuous star formation scenario inferred from the narrow observed M-dwarf separation distributions. This is because it is unclear why stars and BDs should pair differently, avoiding each other while forming in the same way \citep[for a discussion on the observational reality of the BD desert see \S6.3 and \S6.4 in][]{dieterich2012}. Investigating the BD desert for VLM stars together with their separation distribution \emph{in a single sample} could thus lead us closer to answering the question of the (non-)continuity of star formation.

To summarize, the available M-dwarf data do not provide unambigious evidence for star formation that is continuous over and above the hydrogen-burning mass limit. This is because, as illustrated in this paper, the data are also consistent with the here devised DPS model which assumes separate formation modes for stars and BDs and universally valid birth separation distributions for late-type stars. Specifically, the DPS model is \emph{predictive} and able to reproduce both the \minms~and the \al~(with poorer agreement) data, as well as the data for other binary populations observed in a diverse range of environments \citep{mk11,mk12,marks2014,leigh2014bingc}.

\section*{Acknowledgements}
This research was partly supported through DFG grant KR 1635/40-1. The authors thank the referee for providing a thorough report which improved the presentation of the manuscript.

\bibliographystyle{mn2e} \bibliography{biblio} \makeatletter  

\begin{thebibliography}{}

\bibitem[\protect\citeauthoryear{{Abt}}{{Abt}}{1965}]{abt1965a}
{Abt} H.~A.,  1965, \apjs, 11, 429

\bibitem[\protect\citeauthoryear{{Allard}}{{Allard}}{2014}]{allard2014}
{Allard} F.,  2014, in {Booth} M.,  {Matthews} B.~C.,   {Graham} J.~R.,  eds,
  IAU Symposium Vol.~299 of IAU Symposium, {The BT-Settl Model Atmospheres for
  Stars, Brown Dwarfs and Planets}.
pp 271--272

\bibitem[\protect\citeauthoryear{{Banerjee} \& {Kroupa}}{{Banerjee} \&
  {Kroupa}}{2012}]{bk2012r136}
{Banerjee} S.,  {Kroupa} P.,  2012, \aap, 547, A23

\bibitem[\protect\citeauthoryear{{Baraffe} \& {Chabrier}}{{Baraffe} \&
  {Chabrier}}{1996}]{baraffe1996specclass}
{Baraffe} I.,  {Chabrier} G.,  1996, \apjl, 461, L51

\bibitem[\protect\citeauthoryear{{Baraffe}, {Chabrier}, {Barman}, {Allard} \&
  {Hauschildt}}{{Baraffe} et~al.}{2003}]{baraffe2003}
{Baraffe} I.,  {Chabrier} G.,  {Barman} T.~S.,  {Allard} F.,    {Hauschildt}
  P.~H.,  2003, \aap, 402, 701

\bibitem[\protect\citeauthoryear{{Bergfors}, {Brandner}, {Janson}, {Daemgen},
  {Geissler}, {Henning}, {Hippler}, {Hormuth}, {Joergens} \&
  {K{\"o}hler}}{{Bergfors} et~al.}{2010}]{bergfors2010m}
{Bergfors} C.,  {Brandner} W.,  {Janson} M.,  {Daemgen} S.,  {Geissler} K.,
  {Henning} T.,  {Hippler} S.,  {Hormuth} F.,  {Joergens} V.,    {K{\"o}hler}
  R.,  2010, \aap, 520, A54

\bibitem[\protect\citeauthoryear{{Bonnell} \& {Bate}}{{Bonnell} \&
  {Bate}}{2005}]{bb2005}
{Bonnell} I.~A.,  {Bate} M.~R.,  2005, \mnras, 362, 915

\bibitem[\protect\citeauthoryear{{Bouy}, {Brandner}, {Mart{\'{\i}}n},
  {Delfosse}, {Allard} \& {Basri}}{{Bouy} et~al.}{2003}]{bouy2003}
{Bouy} H.,  {Brandner} W.,  {Mart{\'{\i}}n} E.~L.,  {Delfosse} X.,  {Allard}
  F.,    {Basri} G.,  2003, \aj, 126, 1526

\bibitem[\protect\citeauthoryear{{Carquillat} \& {Prieur}}{{Carquillat} \&
  {Prieur}}{2007}]{cp2007a}
{Carquillat} J.-M.,  {Prieur} J.-L.,  2007, \mnras, 380, 1064

\bibitem[\protect\citeauthoryear{{Carrier}, {North}, {Udry} \&
  {Babel}}{{Carrier} et~al.}{2002}]{carrier2002a}
{Carrier} F.,  {North} P.,  {Udry} S.,    {Babel} J.,  2002, \aap, 394, 151

\bibitem[\protect\citeauthoryear{{Clark}, {Blake} \& {Knapp}}{{Clark}
  et~al.}{2012}]{clark2012closemdwarfs}
{Clark} B.~M.,  {Blake} C.~H.,    {Knapp} G.~R.,  2012, \apj, 744, 119

\bibitem[\protect\citeauthoryear{{Close}, {Siegler}, {Freed} \&
  {Biller}}{{Close} et~al.}{2003}]{close2003bd}
{Close} L.~M.,  {Siegler} N.,  {Freed} M.,    {Biller} B.,  2003, \apj, 587,
  407

\bibitem[\protect\citeauthoryear{{Connelley}, {Reipurth} \&
  {Tokunaga}}{{Connelley} et~al.}{2008}]{connelley2008proto}
{Connelley} M.~S.,  {Reipurth} B.,    {Tokunaga} A.~T.,  2008, \aj, 135, 2526

\bibitem[\protect\citeauthoryear{{de la Fuente Marcos} \& {de la Fuente
  Marcos}}{{de la Fuente Marcos} \& {de la Fuente
  Marcos}}{2004}]{fuentemarcos2004}
{de la Fuente Marcos} R.,  {de la Fuente Marcos} C.,  2004, \na, 9, 475

\bibitem[\protect\citeauthoryear{{De Rosa}, {Patience}, {Wilson}, {Schneider},
  {Wiktorowicz}, {Vigan}, {Marois}, {Song}, {Macintosh}, {Graham}, {Doyon},
  {Bessell}, {Thomas} \& {Lai}}{{De Rosa} et~al.}{2014}]{derosa2014vast}
{De Rosa} R.~J.,  {Patience} J.,  {Wilson} P.~A.,  {Schneider} A.,
  {Wiktorowicz} S.~J.,  {Vigan} A.,  {Marois} C.,  {Song} I.,  {Macintosh} B.,
  {Graham} J.~R.,  {Doyon} R.,  {Bessell} M.~S.,  {Thomas} S.,    {Lai} O.,
  2014, \mnras, 437, 1216

\bibitem[\protect\citeauthoryear{{Delfosse}, {Beuzit}, {Marchal}, {Bonfils},
  {Perrier}, {S{\'e}gransan}, {Udry}, {Mayor} \& {Forveille}}{{Delfosse}
  et~al.}{2004}]{d2004m}
{Delfosse} X.,  {Beuzit} J.-L.,  {Marchal} L.,  {Bonfils} X.,  {Perrier} C.,
  {S{\'e}gransan} D.,  {Udry} S.,  {Mayor} M.,    {Forveille} T.,  2004, in
  {Hilditch} R.~W.,  {Hensberge} H.,   {Pavlovski} K.,  eds, Spectroscopically
  and Spatially Resolving the Components of the Close Binary Stars Vol.~318 of
  Astronomical Society of the Pacific Conference Series, {M dwarfs binaries:
  Results from accurate radial velocities and high angular resolution
  observations}.
pp 166--174

\bibitem[\protect\citeauthoryear{{Dhital}, {West}, {Stassun} \&
  {Bochanski}}{{Dhital} et~al.}{2010}]{dhital2010widemdwarfs}
{Dhital} S.,  {West} A.~A.,  {Stassun} K.~G.,    {Bochanski} J.~J.,  2010, \aj,
  139, 2566

\bibitem[\protect\citeauthoryear{{Dieterich}, {Henry}, {Golimowski}, {Krist} \&
  {Tanner}}{{Dieterich} et~al.}{2012}]{dieterich2012}
{Dieterich} S.~B.,  {Henry} T.~J.,  {Golimowski} D.~A.,  {Krist} J.~E.,
  {Tanner} A.~M.,  2012, \aj, 144, 64

\bibitem[\protect\citeauthoryear{{Duch{\^e}ne}}{{Duch{\^e}ne}}{1999}]{d1999}
{Duch{\^e}ne} G.,  1999, \aap, 341, 547

\bibitem[\protect\citeauthoryear{{Duch{\^e}ne} \& {Kraus}}{{Duch{\^e}ne} \&
  {Kraus}}{2013}]{dk2013rev}
{Duch{\^e}ne} G.,  {Kraus} A.,  2013, \araa, 51, 269

\bibitem[\protect\citeauthoryear{Duquennoy \& Mayor}{Duquennoy \&
  Mayor}{1991}]{DuqMay91}
Duquennoy A.,  Mayor M.,  1991, Astronomy and Astrophysics, 248, 485

\bibitem[\protect\citeauthoryear{{Elson}, {Gilmore}, {Santiago} \&
  {Casertano}}{{Elson} et~al.}{1995}]{e1995wcen}
{Elson} R.~A.~W.,  {Gilmore} G.~F.,  {Santiago} B.~X.,    {Casertano} S.,
  1995, \aj, 110, 682

\bibitem[\protect\citeauthoryear{{Fischer} \& {Marcy}}{{Fischer} \&
  {Marcy}}{1992}]{fm1992m}
{Fischer} D.~A.,  {Marcy} G.~W.,  1992, \apj, 396, 178

\bibitem[\protect\citeauthoryear{{Gieles}, {Larsen}, {Bastian} \&
  {Stein}}{{Gieles} et~al.}{2006}]{gieles2006}
{Gieles} M.,  {Larsen} S.~S.,  {Bastian} N.,    {Stein} I.~T.,  2006, \aap,
  450, 129

\bibitem[\protect\citeauthoryear{{Giersz}, {Heggie}, {Hurley} \&
  {Hypki}}{{Giersz} et~al.}{2013}]{giersz2013mocca}
{Giersz} M.,  {Heggie} D.~C.,  {Hurley} J.~R.,    {Hypki} A.,  2013, \mnras,
  431, 2184

\bibitem[\protect\citeauthoryear{{Goodwin} \& {Kroupa}}{{Goodwin} \&
  {Kroupa}}{2005}]{gk2005}
{Goodwin} S.~P.,  {Kroupa} P.,  2005, \aap, 439, 565

\bibitem[\protect\citeauthoryear{{Grether} \& {Lineweaver}}{{Grether} \&
  {Lineweaver}}{2006}]{grether2006desert}
{Grether} D.,  {Lineweaver} C.~H.,  2006, \apj, 640, 1051

\bibitem[\protect\citeauthoryear{{Hauschildt}, {Allard}, {Ferguson}, {Baron} \&
  {Alexander}}{{Hauschildt} et~al.}{1999}]{hauschildt1999}
{Hauschildt} P.~H.,  {Allard} F.,  {Ferguson} J.,  {Baron} E.,    {Alexander}
  D.~R.,  1999, \apj, 525, 871

\bibitem[\protect\citeauthoryear{{Janson}, {Bergfors}, {Brandner},
  {Kudryavtseva}, {Hormuth}, {Hippler} \& {Henning}}{{Janson}
  et~al.}{2014}]{janson2014m}
{Janson} M.,  {Bergfors} C.,  {Brandner} W.,  {Kudryavtseva} N.,  {Hormuth} F.,
   {Hippler} S.,    {Henning} T.,  2014, \apj, 789, 102

\bibitem[\protect\citeauthoryear{{Janson}, {Hormuth}, {Bergfors}, {Brandner},
  {Hippler}, {Daemgen}, {Kudryavtseva}, {Schmalzl}, {Schnupp} \&
  {Henning}}{{Janson} et~al.}{2012}]{janson2012m}
{Janson} M.,  {Hormuth} F.,  {Bergfors} C.,  {Brandner} W.,  {Hippler} S.,
  {Daemgen} S.,  {Kudryavtseva} N.,  {Schmalzl} E.,  {Schnupp} C.,    {Henning}
  T.,  2012, \apj, 754, 44

\bibitem[\protect\citeauthoryear{{King}, {Goodwin}, {Parker} \&
  {Patience}}{{King} et~al.}{2012}]{King2012b}
{King} R.~R.,  {Goodwin} S.~P.,  {Parker} R.~J.,    {Patience} J.,  2012,
  \mnras, 427, 2636

\bibitem[\protect\citeauthoryear{{Kohler} \& {Leinert}}{{Kohler} \&
  {Leinert}}{1998}]{kl1998}
{Kohler} R.,  {Leinert} C.,  1998, \aap, 331, 977

\bibitem[\protect\citeauthoryear{{K{\"o}hler}, {Neuh{\"a}user}, {Kr{\"a}mer},
  {Leinert}, {Ott} \& {Eckart}}{{K{\"o}hler} et~al.}{2008}]{kn2008cra}
{K{\"o}hler} R.,  {Neuh{\"a}user} R.,  {Kr{\"a}mer} S.,  {Leinert} C.,  {Ott}
  T.,    {Eckart} A.,  2008, \aap, 488, 997

\bibitem[\protect\citeauthoryear{{Korntreff}, {Kaczmarek} \&
  {Pfalzner}}{{Korntreff} et~al.}{2012}]{kkp2012}
{Korntreff} C.,  {Kaczmarek} T.,    {Pfalzner} S.,  2012, \aap, 543, A126

\bibitem[\protect\citeauthoryear{{Kraus}, {Ireland}, {Martinache} \&
  {Hillenbrand}}{{Kraus} et~al.}{2011}]{kraus2011tau}
{Kraus} A.~L.,  {Ireland} M.~J.,  {Martinache} F.,    {Hillenbrand} L.~A.,
  2011, \apj, 731, 8

\bibitem[\protect\citeauthoryear{{Kroupa}}{{Kroupa}}{1995}]{k95b}
{Kroupa} P.,  1995, \mnras, 277, 1507

\bibitem[\protect\citeauthoryear{{Kroupa}}{{Kroupa}}{2001}]{k2001}
{Kroupa} P.,  2001, \mnras, 322, 231

\bibitem[\protect\citeauthoryear{{Kroupa}}{{Kroupa}}{2011}]{k2011}
{Kroupa} P.,  2011, in {Alves} J.,  {Elmegreen} B.~G.,  {Girart} J.~M.,
  {Trimble} V.,  eds, Computational Star Formation Vol.~270 of IAU Symposium,
  {The universality hypothesis: binary and stellar populations in star clusters
  and galaxies}.
pp 141--149

\bibitem[\protect\citeauthoryear{{Kroupa} \& {Bouvier}}{{Kroupa} \&
  {Bouvier}}{2003}]{kb2003tau}
{Kroupa} P.,  {Bouvier} J.,  2003, \mnras, 346, 343

\bibitem[\protect\citeauthoryear{{Kroupa}, {Bouvier}, {Duch{\^e}ne} \&
  {Moraux}}{{Kroupa} et~al.}{2003}]{kroupa2003bd}
{Kroupa} P.,  {Bouvier} J.,  {Duch{\^e}ne} G.,    {Moraux} E.,  2003, \mnras,
  346, 354

\bibitem[\protect\citeauthoryear{{Kroupa} \& {Petr-Gotzens}}{{Kroupa} \&
  {Petr-Gotzens}}{2011}]{kp2011}
{Kroupa} P.,  {Petr-Gotzens} M.~G.,  2011, \aap, 529, A92

\bibitem[\protect\citeauthoryear{{Kroupa}, {Weidner}, {Pflamm-Altenburg},
  {Thies}, {Dabringhausen}, {Marks} \& {Maschberger}}{{Kroupa}
  et~al.}{2013}]{k2013rev}
{Kroupa} P.,  {Weidner} C.,  {Pflamm-Altenburg} J.,  {Thies} I.,
  {Dabringhausen} J.,  {Marks} M.,    {Maschberger} T.,  2013, {The Stellar and
  Sub-Stellar Initial Mass Function of Simple and Composite Populations}.
p.~115

\bibitem[\protect\citeauthoryear{{Lada} \& {Lada}}{{Lada} \&
  {Lada}}{2003}]{ll2003}
{Lada} C.~J.,  {Lada} E.~A.,  2003, \araa, 41, 57

\bibitem[\protect\citeauthoryear{{Leigh}, {Giersz}, {Webb}, {Hypki}, {De
  Marchi}, {Kroupa} \& {Sills}}{{Leigh} et~al.}{2013}]{leigh2013}
{Leigh} N.,  {Giersz} M.,  {Webb} J.~J.,  {Hypki} A.,  {De Marchi} G.,
  {Kroupa} P.,    {Sills} A.,  2013, \mnras, 436, 3399

\bibitem[\protect\citeauthoryear{{Leigh}, {Umbreit}, {Sills}, {Knigge}, {de
  Marchi}, {Glebbeek} \& {Sarajedini}}{{Leigh} et~al.}{2012}]{leigh2012}
{Leigh} N.,  {Umbreit} S.,  {Sills} A.,  {Knigge} C.,  {de Marchi} G.,
  {Glebbeek} E.,    {Sarajedini} A.,  2012, \mnras, 422, 1592

\bibitem[\protect\citeauthoryear{{Leigh}, {Giersz}, {Marks}, {Webb}, {Hypki},
  {Heinke}, {Kroupa} \& {Sills}}{{Leigh} et~al.}{2014}]{leigh2014bingc}
{Leigh} N.~W.~C.,  {Giersz} M.,  {Marks} M.,  {Webb} J.~J.,  {Hypki} A.,
  {Heinke} C.~O.,  {Kroupa} P.,    {Sills} A.,  2014, ArXiv e-prints

\bibitem[\protect\citeauthoryear{{Leinert}, {Zinnecker}, {Weitzel}, {Christou},
  {Ridgway}, {Jameson}, {Haas} \& {Lenzen}}{{Leinert} et~al.}{1993}]{lz1993tau}
{Leinert} C.,  {Zinnecker} H.,  {Weitzel} N.,  {Christou} J.,  {Ridgway} S.~T.,
   {Jameson} R.,  {Haas} M.,    {Lenzen} R.,  1993, \aap, 278, 129

\bibitem[\protect\citeauthoryear{{Li}, {Kouwenhoven}, {Stamatellos} \&
  {Goodwin}}{{Li} et~al.}{2015}]{li2015fragmentation}
{Li} Y.,  {Kouwenhoven} M.~B.~N.,  {Stamatellos} D.,    {Goodwin} S.~P.,  2015,
  \apj, 805, 116

\bibitem[\protect\citeauthoryear{Marks \& Kroupa}{Marks \& Kroupa}{2011}]{mk11}
Marks M.,  Kroupa P.,  2011, Monthly Notices of the Royal Astronomical Society,
  417, 1702

\bibitem[\protect\citeauthoryear{Marks \& Kroupa}{Marks \& Kroupa}{2012}]{mk12}
Marks M.,  Kroupa P.,  2012, Astronomy \& Astrophysics, 543, 14

\bibitem[\protect\citeauthoryear{Marks, Kroupa \& Oh}{Marks
  et~al.}{2011}]{mko11}
Marks M.,  Kroupa P.,    Oh S.,  2011, Monthly Notices of the Royal
  Astronomical Society, 417, 1684

\bibitem[\protect\citeauthoryear{{Marks}, {Leigh}, {Giersz}, {Pfalzner},
  {Pflamm-Altenburg} \& {Oh}}{{Marks} et~al.}{2014}]{marks2014}
{Marks} M.,  {Leigh} N.,  {Giersz} M.,  {Pfalzner} S.,  {Pflamm-Altenburg} J.,
    {Oh} S.,  2014, \mnras, 441, 3503

\bibitem[\protect\citeauthoryear{{Mayor}, {Duquennoy}, {Halbwachs} \&
  {Mermilliod}}{{Mayor} et~al.}{1992}]{mayor1992k}
{Mayor} M.,  {Duquennoy} A.,  {Halbwachs} J.-L.,    {Mermilliod} J.-C.,  1992,
  in {McAlister} H.~A.,  {Hartkopf} W.~I.,  eds, IAU Colloq. 135: Complementary
  Approaches to Double and Multiple Star Research Vol.~32 of Astronomical
  Society of the Pacific Conference Series, {CORAVEL Surveys to Study Binaries
  of Different Masses and Ages}.
p.~73

\bibitem[\protect\citeauthoryear{{McCarthy}, {Zuckerman} \&
  {Becklin}}{{McCarthy} et~al.}{2003}]{mzb2003desert}
{McCarthy} C.,  {Zuckerman} B.,    {Becklin} E.~E.,  2003, in {Mart{\'{\i}}n}
  E.,  ed., Brown Dwarfs Vol.~211 of IAU Symposium, {There is a Brown Dwarf
  Desert of Companions Orbiting Stars between 75 and 1000 AU}.
p.~279

\bibitem[\protect\citeauthoryear{{Milone}, {Piotto}, {Bedin}, {Aparicio} \&
  {Anderson}}{{Milone} et~al.}{2012}]{m2012acs}
{Milone} A.~P.,  {Piotto} G.,  {Bedin} L.~R.,  {Aparicio} A.,    {Anderson} J.,
   2012, \aap, 540, A16

\bibitem[\protect\citeauthoryear{{Oh}, {Kroupa} \& {Pflamm-Altenburg}}{{Oh}
  et~al.}{2015}]{okp2015}
{Oh} S.,  {Kroupa} P.,    {Pflamm-Altenburg} J.,  2015, ApJ, in press

\bibitem[\protect\citeauthoryear{{Parker} \& {Meyer}}{{Parker} \&
  {Meyer}}{2014}]{pm2014}
{Parker} M.,  {Meyer} M.,  2014, \mnras

\bibitem[\protect\citeauthoryear{{Parker}}{{Parker}}{2014}]{parker2014}
{Parker} R.~J.,  2014, ArXiv e-prints

\bibitem[\protect\citeauthoryear{{Parker}, {Goodwin} \& {Allison}}{{Parker}
  et~al.}{2011}]{pga2011}
{Parker} R.~J.,  {Goodwin} S.~P.,    {Allison} R.~J.,  2011, \mnras, 418, 2565

\bibitem[\protect\citeauthoryear{{Parker} \& {Reggiani}}{{Parker} \&
  {Reggiani}}{2013}]{parker2013mr}
{Parker} R.~J.,  {Reggiani} M.~M.,  2013, \mnras, 432, 2378

\bibitem[\protect\citeauthoryear{{Raghavan}, {McAlister}, {Henry}, {Latham},
  {Marcy}, {Mason}, {Gies}, {White} \& {ten Brummelaar}}{{Raghavan}
  et~al.}{2010}]{r2010}
{Raghavan} D.,  {McAlister} H.~A.,  {Henry} T.~J.,  {Latham} D.~W.,  {Marcy}
  G.~W.,  {Mason} B.~D.,  {Gies} D.~R.,  {White} R.~J.,    {ten Brummelaar}
  T.~A.,  2010, \apjs, 190, 1

\bibitem[\protect\citeauthoryear{{Reggiani} \& {Meyer}}{{Reggiani} \&
  {Meyer}}{2011}]{regmey2011}
{Reggiani} M.~M.,  {Meyer} M.~R.,  2011, \apj, 738, 60

\bibitem[\protect\citeauthoryear{{Reipurth}, {Clarke}, {Boss}, {Goodwin},
  {Rodriguez}, {Stassun}, {Tokovinin} \& {Zinnecker}}{{Reipurth}
  et~al.}{2014}]{reipurth2014rev}
{Reipurth} B.,  {Clarke} C.~J.,  {Boss} A.~P.,  {Goodwin} S.~P.,  {Rodriguez}
  L.~F.,  {Stassun} K.~G.,  {Tokovinin} A.,    {Zinnecker} H.,  2014, ArXiv
  e-prints

\bibitem[\protect\citeauthoryear{{Sana}, {de Mink}, {de Koter}, {Langer},
  {Evans}, {Gieles}, {Gosset}, {Izzard}, {Le Bouquin} \& {Schneider}}{{Sana}
  et~al.}{2012}]{sana2012o}
{Sana} H.,  {de Mink} S.~E.,  {de Koter} A.,  {Langer} N.,  {Evans} C.~J.,
  {Gieles} M.,  {Gosset} E.,  {Izzard} R.~G.,  {Le Bouquin} J.-B.,
  {Schneider} F.~R.~N.,  2012, Science, 337, 444

\bibitem[\protect\citeauthoryear{{Stahler}}{{Stahler}}{2010}]{stahler2010}
{Stahler} S.~W.,  2010, \mnras, 402, 1758

\bibitem[\protect\citeauthoryear{{Stamatellos} \& {Whitworth}}{{Stamatellos} \&
  {Whitworth}}{2009}]{stamwhit2009discfrag}
{Stamatellos} D.,  {Whitworth} A.~P.,  2009, \mnras, 392, 413

\bibitem[\protect\citeauthoryear{{Thies} \& {Kroupa}}{{Thies} \&
  {Kroupa}}{2007}]{tk2007}
{Thies} I.,  {Kroupa} P.,  2007, \apj, 671, 767

\bibitem[\protect\citeauthoryear{{Thies} \& {Kroupa}}{{Thies} \&
  {Kroupa}}{2008}]{tk2008tau}
{Thies} I.,  {Kroupa} P.,  2008, \mnras, 390, 1200

\bibitem[\protect\citeauthoryear{{Thies}, {Kroupa}, {Goodwin}, {Stamatellos} \&
  {Whitworth}}{{Thies} et~al.}{2010}]{thies2010bd}
{Thies} I.,  {Kroupa} P.,  {Goodwin} S.~P.,  {Stamatellos} D.,    {Whitworth}
  A.~P.,  2010, \apj, 717, 577

\bibitem[\protect\citeauthoryear{{Thies}, {Pflamm-Altenburg}, {Kroupa} \&
  {Marks}}{{Thies} et~al.}{2015}]{thies2015}
{Thies} I.,  {Pflamm-Altenburg} J.,  {Kroupa} P.,    {Marks} M.,  2015, ArXiv
  e-prints

\bibitem[\protect\citeauthoryear{{Tokovinin}}{{Tokovinin}}{2014}]{tok2014fg}
{Tokovinin} A.,  2014, \aj, 147, 87

\bibitem[\protect\citeauthoryear{{van Leeuwen}}{{van
  Leeuwen}}{2007}]{vanleuwen2007}
{van Leeuwen} F.,  2007, \aap, 474, 653

\bibitem[\protect\citeauthoryear{{Ward-Duong}, {Patience}, {De Rosa}, {Bulger},
  {Rajan}, {Goodwin}, {Parker}, {McCarthy} \& {Kulesa}}{{Ward-Duong}
  et~al.}{2015}]{wardduong2015}
{Ward-Duong} K.,  {Patience} J.,  {De Rosa} R.~J.,  {Bulger} J.,  {Rajan} A.,
  {Goodwin} S.~P.,  {Parker} R.~J.,  {McCarthy} D.~W.,    {Kulesa} C.,  2015,
  \mnras, 449, 2618

\end{thebibliography}
\renewcommand{\@biblabel}[1]{[#1]}   \makeatother

\label{lastpage}

\end{document}